# Dynamics and stationarity of two coupled arbitrary oscillators interacting with separate reservoirs

Illarion Dorofeyev[*]

Institute for Physics of Microstructures, Russian Academy of Sciences,

603950, GSP-105 Nizhny Novgorod, Russia

## Abstract

This work addresses the problem of relaxation of open systems to quasi-equilibrium states. Time-dependent density matrix of two arbitrary coupled quantum oscillators of arbitrary properties interacting with separate reservoirs is derived based on path integration. Temporal dynamics of spatial variances and covariances of the oscillators from any given time up to quasi-equilibrium steady states is studied. It is demonstrated for general case that asymptotic spatial variances of two arbitrary oscillators and their covariances achieve stationary values in the long-time limit. A comparison of steady state characteristics of coupled oscillators with those predicted by the fluctuation dissipation theorem (FDT) is performed. It is shown that the larger the difference in masses and eigenfrequencies of coupled oscillators, the smaller are the deviations of stationary variances from those given by the FDT at fixed coupling strength and fixed difference in temperatures between thermal baths. In framework of the model of the bilinear interaction and Ohmic dissipation and at any oscillators' parameters the variances and covariances have divergent character at strong couplings.

*) E-mail: Illarion1955@mail.ru

# I. Introduction

Physical phenomena of interaction between bodies kept with different temperatures are very important in nature. The difference in temperatures stimulates new effects within systems out of equilibrium as it was established previously, for example in [1-12]. We underline that in all cited papers the stationary systems out of equilibrium were studied. Obviously, that the temporal dynamics of interacting systems out of equilibrium and their possible steady states take special studies. A quantum system interacting with a background is often called as the open quantum system. The physics of such systems always attracts a great interest in science and technology. For an appropriate model of the open quantum system it is necessary to describe most essential aspects of the dynamical behavior of the system under study and possible stationarity. As a result of this, an oscillator or set of oscillators coupled to reservoirs of independent oscillators are generally accepted models of open quantum systems. For example, these models have been successful in describing the Brownian dynamics of selected particles coupled to a bath of harmonic oscillators, see [13-25]. Among methods to study above mentioned problems the path integral technique has a special role due to universality [22, 26, 27]. The path integration allows for a complete solution to this problem both for quantum and classical limits. Irreversibility in the Brownian dynamics of selected particle interacting with a bath appears after reduction over the reservoir's variables. In such a way, an explicit expression for the time evolution of the density matrix of an oscillator was obtained and investigated in detail in [28]. The spatial dispersion of the oscillator reaches a steady state and fully agrees with the fluctuation-disspation theorem (FDT). A transition to some intermediate stationary states of oscillators interacting with heat baths in different approaches was studied in [29-35]. It was shown, for instance, that an arbitrary initial state of a harmonic oscillator relaxes towards a uniquely determined stationary state. Taking into account the problem of our study, we concentrate our attention upon the papers describing bipartite systems composed of two oscillators. Different stages of entanglement evolution and quantum discord between two oscillators coupled to a common environment was studied in [36-38] for different models of environments in the non-Markovian regime. Using non-Markovian master equations the entanglement evolution of two harmonic oscillators under the influence of non-Markovian thermal environments was studied in [39]. It was shown that the dynamics of the quantum entanglement is sensitive to the initial states, the coupling between oscillators and the coupling to a common bath or to independent baths.

Quantum decoherence of two coupled oscillators in a common environment at arbitrary temperature was investigated in [40] and was shown that the problem can be mapped into that of a single harmonic oscillator in a general environment plus a free harmonic oscillator. Properties of the stationary state of a bipartite system may differ. In [41] the temporal evolution of quantum correlations of entangled two-mode states was examined and found that in the two-reservoir model the initial entanglement is completely lost, and both modes are finally uncorrelated, but in a common reservoir both modes interact indirectly via the same bath. The temporal behavior of entanglement between two coupled nonidentical oscillators in contact with a common bath and with two separate baths was studied in [42] based on a full master equation. It was shown that the initial two-mode squeezed state becomes separable in case of separate baths at some conditions, but in case of a common bath, the asymptotic entanglement becomes possible. A system of two coupled oscillators interacting with independent reservoirs was investigated in [43] and was shown that the system does not have a stationary state for all linear interactions. The interaction between subsystems must be strong to be in a steady state entanglement. No thermal entanglement is found in the high-temperature regime and weak coupling limits in [44]. A steady state regime of two coupled oscillators within independent heat reservoirs of harmonic oscillators was investigated in [45], and an analytical expression for their mean energy of interaction was derived. Temporal dynamics of variances and covariances in the weak-coupling limit was studied in [46]. It was shown that the system of two weakly coupled oscillators in the infinite time limit agrees with the fluctuation dissipation theorem despite on initial conditions. The case of arbitrary coupling of identical oscillators was considered in [47] and shown that the larger a difference in temperatures of thermal baths, the larger is a difference of the stationary values of variances of coupled identical oscillators as compared to values given by the fluctuation dissipation theorem.

The paper addresses the time-dependent dynamics of two arbitrary coupled oscillators of arbitrary properties interacting with independent heat reservoirs kept, in general, at different temperatures. We show and analyze the reachability of quasi-equilibrium stationary states from given initial states of the system of two arbitrary oscillators interacting with separate baths at different coupling strengths and at different temperatures. This work is the generalization of a research in [46, 47] devoted to the partial problems of the relaxation dynamics of weakly-coupled different oscillators and of arbitrary coupled identical oscillators.

The paper is organized as follows. In Sec.II we provide an expression for the total and reduced time-dependent density matrix of two arbitrary coupled quantum oscillators of arbitrary properties interacting with different thermal baths and a brief description of obtaining the density matrix based on the path integration method. A comparison of a general case with more special cases corresponding to the weak-coupled oscillators and to identical oscillators, and also numerical study of a temporal behavior of variances and covariances from given initial states up to states in the infinite time limit are given in Sec.III. Final conclusions are given in Sec.IV.

## II. Problem statement and solution

We consider the system of two coupled oscillators where each oscillator is connected to a separate reservoir of independent harmonic oscillators. The Hamiltonian of the system is given by

$$H = p_1^2/2M_1 + M_1\omega_{01}^2 x_1^2/2 + p_2^2/2M_2 + M_2\omega_{02}^2 x_2^2/2 - \lambda x_1 x_2 + \\ + \sum_{j=1}^{N_1}\left[p_j^2/2m_j + m_j\omega_j^2(q_j - x_1)^2/2\right] + \sum_{k=1}^{N_2}\left[p_k^2/2m_k + m_k\omega_k^2(q_k - x_2)^2/2\right], \quad (1)$$

where $x_{1,2}, p_{1,2}, M_{1,2}, \omega_{01,02}$ are the coordinates, momenta, masses and eigenfrequencies of the selected oscillators, $\pm\lambda$ is the coupling constant, $q_j, p_j, \omega_j, m_j$ and $q_k, p_k, \omega_k, m_k$ are the coordinates, momenta, eigenfrequencies and masses of bath's oscillators. Both selected oscillators have arbitrary masses $M_{1,2}$, eigenfrequencies $\omega_{01,02}$ and their coupling constant ranging in value $|\lambda| \leq \omega_{01}\omega_{02}\sqrt{M_1 M_2}$ for stability. The numbers of modes $N_{1,2}$ composing the reservoirs must be large enough in order to make up the Poincare circle to be time longer than any relaxation processes within the system under study. In time $t < 0$ all subsystems of oscillators are uncoupled and the density matrix of the whole system is factorized. Then, the interactions are switched on in the time $t = 0$ and maintained during arbitrary time interval up to infinity. The problem is to find the time-dependent density matrix of two arbitrary coupled oscillators of arbitrary parameters in any moment of time $t \geq 0$. We follow the prescription given in [28, 49] to obtain a reduced density matrix in general case as well as in case of the weak coupling limit [46] and in a special case of identical oscillators [47]. Here we describe the key aspects of this method, referring the reader for further details to the cited publication. The density matrix $W(t)$ of the total Hamiltonian system at any time $t$ can be obtained by the unitary transformation as follows

$$W(t) \equiv W_t = \exp(-iHt/\hbar)W(0)\exp(iHt/\hbar), \quad (2)$$

where $W(0)$ is the initial density matrix of the total isolated system at $t = 0$, which is chosen in the factorized form

$$W(t=0) \equiv W_0 = \rho_A^{(1)}(0)\rho_A^{(2)}(0)\rho_B^{(1)}(0)\rho_B^{(2)}(0), \quad (3)$$

where $\rho_A^{(1),(2)}(0)$ are the initially prepared density matrices of the two selected oscillators, $\rho_B^{(1),(2)}(0)$ are initial density matrices of the separate reservoirs. Then we use the completeness properties of the eigenfunctions in the coordinate representation

$$\int dx_1 dx_2 d\vec{R}_1 d\vec{R}_2 \,|x_1 x_2 \vec{R}_1 \vec{R}_2\rangle\langle x_1 x_2 \vec{R}_1 \vec{R}_2| \equiv \int d\vec{x}d\vec{R}\,|\vec{x}\vec{R}\rangle\langle \vec{x}\vec{R}| = 1, \qquad (4)$$

where the limits of the multiple integration extended from minus to plus infinity and we use the designations $\vec{R}_1 = \{q_j\} = \{q_1,...,q_{N_1}\}$ and $\vec{R}_2 = \{q_k\} = \{q_1,...,q_{N_2}\}$, $\vec{x} = \{x_1, x_2\}$ and $\vec{R} = \{\vec{R}_1, \vec{R}_2\}$.

Using Eq.(4), the Eq. (2) can be written in the matrix form as follows

$$\langle \vec{x}\vec{R}|W_t|\vec{y}\vec{Q}\rangle = \int d\vec{x}'d\vec{R}'d\vec{y}'d\vec{Q}' \langle \vec{x}\vec{R}|e^{-iHt/\hbar}|\vec{x}'\vec{R}'\rangle\langle \vec{x}'\vec{R}'|W_0|\vec{y}'\vec{Q}'\rangle\langle \vec{y}'\vec{Q}'|e^{iHt/\hbar}|\vec{y}\vec{Q}\rangle, \qquad (5)$$

The transition amplitudes in Eq.(5) are expressed via the path integrals [48-51], for example

$$\langle \vec{x}\vec{R}|e^{-iHt/\hbar}|\vec{x}'\vec{R}'\rangle = \int \mathsf{D}x_1 \mathsf{D}x_2 \mathsf{D}\vec{R}_1 \mathsf{D}\vec{R}_2 \exp\{(i/\hbar)S[x_1(\tau), x_2(\tau), \vec{R}_1(\tau), \vec{R}_2(\tau)]\}, \qquad (6)$$

where the integration along all paths is carried out from $x_1(0) = x_1'$ to $x_1(t) = x_1$, from $x_2(0) = x_2'$ to $x_2(t) = x_2$, and from $\vec{R}_1(0) = \vec{R}_1'$ to $\vec{R}_1(t) = \vec{R}_1$, from $\vec{R}_2(0) = \vec{R}_2'$ to $\vec{R}_2(t) = \vec{R}_2$. The measures $\mathsf{D}x_i \mathsf{D}\vec{R}_i, (i=1,2)$ are expressed via the multiple Reihmann integrals in the coordinate space as usual [48]. The action $S[x_1, x_2, \vec{R}_1, \vec{R}_2]$ in Eq.(6) is expressed via the Lagrangian corresponding to the Hamiltonian in Eq.(1).

Then, after reduction of the total density matrix in Eq.(5) over the coordinates of all baths oscillators followed by calculation of the Feynman-Vernon influence functional for our case we obtained the reduced density matrix as well as in [46]. This density matrix can be written as follows

$$\rho(x_1, x_2, y_1, y_2, t) = \int dx_1' dx_2' dy_1' dy_2' \, J(x_1, x_2, y_1, y_2, t; x_1', x_2', y_1', y_2', 0) \, \rho_A^{(1)}(x_1', y_1', 0)\rho_A^{(2)}(x_2', y_2', 0), \qquad (7)$$

where $J(x_1, x_2, y_1, y_2, t; x_1', x_2', y_1', y_2', 0)$ is the propagator calculated for this problem in [46].

The final expression is obtained in new variables $X_{1,2} = x_{1,2} + y_{1,2}$, $\xi_{1,2} = x_{1,2} - y_{1,2}$. In these new variables the density matrix in Eq. (7) reads

$$\begin{aligned}\rho(X_{f1}, X_{f2}, \xi_{f1}, \xi_{f2}, t) = \int dX_{i1} dX_{i2} d\xi_{i1} d\xi_{i2} \, &J(X_{f1}, X_{f2}, \xi_{f1}, \xi_{f2}, t; X_{i1}, X_{i2}, \xi_{i1}, \xi_{i2}, 0) \\ &\times \rho_A^{(1)}(X_{i1}, \xi_{i1}, 0)\rho_A^{(2)}(X_{i2}, \xi_{i2}, 0)\end{aligned}, \qquad (8)$$

where $X_{i1,i2} = X_{1,2}(0)$, $X_{f1,f2} = X_{1,2}(t)$ and $\xi_{i1,i2} = \xi_{1,2}(0)$, $\xi_{f1,f2} = \xi_{1,2}(t)$, and the propagating function in Eq.(8) can be represented in the following form

$$J(X_{f1}, X_{f2}, \xi_{f1}, \xi_{f2}, t; X_{i1}, X_{i2}, \xi_{i1}, \xi_{i2}, 0) =$$

$$\tilde{C}_1 \tilde{C}_2 F_1^2(t) F_2^2(t) \exp\frac{i}{\hbar}\{\tilde{S}_{cl}^{(1)}(t) + \tilde{S}_{cl}^{(2)}(t) + \tilde{S}_{cl}^{(12)}(t)\}$$

$$\times \exp-\frac{1}{\hbar}\{A_1(t)\xi_{f1}^2 + B_1(t)\xi_{f1}\xi_{i1} + C_1(t)\xi_{i1}^2\} \qquad (9)$$

$$\times \exp-\frac{1}{\hbar}\{A_2(t)\xi_{f2}^2 + B_2(t)\xi_{f2}\xi_{i2} + C_2(t)\xi_{i2}^2\}$$

$$\times \exp-\frac{1}{\hbar}\{E_1(t)\xi_{i1}\xi_{i2} + E_2(t)\xi_{f2}\xi_{i1} + E_3(t)\xi_{f1}\xi_{i2} + E_4(t)\xi_{f1}\xi_{f2}\}$$

where expressions for the classical actions $\tilde{S}_{cl}^{(1)}$, $\tilde{S}_{cl}^{(2)}$ and $\tilde{S}_{cl}^{(12)}$ are formally the same as in [46], but with other all functions involved. For the general case of arbitrary coupled oscillators of arbitrary properties interacting with baths kept at different temperatures all the time-dependent functions from Eq.(9) are presented in Appendix A. The tilded classical actions appear in Eq.(9) because we represent the paths as the sums $X_{1,2} = \tilde{X}_{1,2} + X'_{1,2}$, $\xi_{1,2} = \tilde{\xi}_{1,2} + \xi'_{1,2}$, explicitly selecting classical paths $\tilde{X}_{1,2}$, $\tilde{\xi}_{1,2}$ and fluctuating parts $X'_{1,2}$, $\xi'_{1,2}$ with boundary conditions $X'_{1,2}(0) = X'_{1,2}(t) = 0$, $\xi'_{1,2}(0) = \xi'_{1,2}(t) = 0$. Evaluation of the related fluctuational integral describing fluctuations $X'_{1,2}$ and $\xi'_{1,2}$ of paths around the classical paths $\tilde{X}_{1,2}$ and $\tilde{\xi}_{1,2}$ in a general case was done in Appendix D in [46]. This fluctuational part gives the prefactor in Eq.(9). The classical problem for general case of arbitrary coupled oscillators is solved in Appendix B of this paper.

Initial density matrices of each of two oscillators in Eq.(8) are chosen in the Gaussian forms as follows

$$\rho_A^{(k)}(X_{ik}, \xi_{ik}, 0) = (2\pi\sigma_{0k}^2)^{-1/2} \exp\left[-(X_{ik}^2 + \xi_{ik}^2)/8\sigma_{0k}^2\right], \quad (k = 1, 2) \qquad (10)$$

where $\sigma_{01}^2$ and $\sigma_{02}^2$ are the initial dispersions of oscillators.

### III. Results and discussion.

#### A. The density matrix for coupled arbitrary oscillators out of equilibrium.

After an integration in Eq.(8) with use of Eqs.(9), (10) the density matrix of coupled oscillators of arbitrary properties and arbitrary coupling strength is obtained in the following form

$$\begin{aligned}
\rho(X_{f1}, X_{f2}, \xi_{f1}, \xi_{f2}, t) &= \rho_0(t) \\
&\times \exp\left\{-\left[g_1(t)X_{f1}^2 + g_{12}(t)X_{f1}X_{f2} + g_2(t)X_{f2}^2\right]\right\} \\
&\times \exp\left\{-\left[g_1'(t)\xi_{f1}^2 + g_{12}'(t)\xi_{f1}\xi_{f2} + g_2'(t)\xi_{f2}^2\right]\right\} \\
&\times \exp\left\{-\left[g_{11}''(t)X_{f1}\xi_{f1} + g_{21}''(t)X_{f2}\xi_{f1} + g_{12}''(t)X_{f1}\xi_{f2} + g_{22}''(t)X_{f2}\xi_{f2}\right]\right\}
\end{aligned} \qquad (11)$$

where all functions $\rho_0(t)$, $g(t)$, $g'(t)$ and $g''(t)$ are written in Appendix C.

For simplicity as well as in [28] we put $X_{fk} = 2x_{fk}$ and $\xi_{fk} = 0$, $(k=1,2)$ in Eq.(11) to transform this expression into the compact form

$$\rho(x_{f1}, x_{f2}, t) = \rho_0(t) \exp\left[-\frac{1}{2}\beta_{11}(t)x_{f1}^2 - \beta_{12}(t)x_{f1}x_{f2} - \frac{1}{2}\beta_{22}(t)x_{f2}^2\right], \qquad (12)$$

where

$$\begin{aligned}
\beta_{11}(t,\lambda,T_1,T_2) &= 8g_1(t,\lambda,T_1,T_2), \quad \beta_{22}(t,\lambda,T_1,T_2) = 8g_2(t,\lambda,T_1,T_2), \\
\beta_{12}(t,\lambda,T_1,T_2) &= 8g_{12}(t,\lambda,T_1,T_2),
\end{aligned} \qquad (13)$$

where $g_1$, $g_2$, $g_{12}$ are listed in Eqs.(C2) in Appendix C.

From Eqs.(12), (13) we obtain corresponding second moments as prescribed [52], which are written here in the short form

$$\sigma_1^2(t) \equiv \langle x_{f1}^2 \rangle_t = \frac{\beta_{22}(t)}{\beta_{11}(t)\beta_{22}(t) - \beta_{12}^2(t)}, \quad \sigma_2^2(t) \equiv \langle x_{f2}^2 \rangle_t = \frac{\beta_{11}(t)}{\beta_{11}(t)\beta_{22}(t) - \beta_{12}^2(t)}, \qquad (14)$$

$$\beta_{12}^{-1}(t) \equiv \langle x_{f1}x_{f2} \rangle_t = \frac{\beta_{12}(t)}{\beta_{12}^2(t) - \beta_{11}(t)\beta_{22}(t)}. \qquad (15)$$

We note that in general case the variances and covariance in Eqs.(14),(15) depend on both temperatures $T_1$ and $T_2$, on arbitrary different parameters of oscillators and on arbitrary coupling strength between oscillators.

We note here that, from obtained formulas related to a general case of arbitrary coupled oscillators of arbitrary properties, all obtained results follow for the cases of weakly-coupled oscillators, arbitrary coupled identical oscillators in [46, 47] and for a single oscillator interacting with a thermal bath [28].

### B. Time-dependent dynamics of spatial variances and steady states of coupled oscillators of arbitrary properties.

In this paragraph we begin with a study of steady states of two coupled arbitrary oscillators interacting with separate baths kept at fifferent temperatures. Qualitatevely the states can be obtained by the asymptotic transition $t \to \infty$ in all formulas. Analytically, this can be done in case of weakly coupling between the oscillators due to relative simplicity [46]. In common case, it is practically impossible because of complexity of time-dependent relaxation dynamics during the initial phase of the Poincare circle of Hamiltoinian system. In our numerical calculations we identified the steady states by putting $v_m t \gg 1$, where $v_m = Min\{\gamma, \Omega_{1,2}, \omega_{01,02}\}$.

It can be seen from Eq.(9) for the propagator that the classical paths play one of the crucial role in dynamics of the system under study. In a general case of coupled oscillators of arbitrary parameters the coefficients $r_1$ and $r_2$ are very important characteristics, see [53-55] and Appendix B, because they show specific contributions from the first and second modes to the classical trajectories of coupled oscillators. The case of identical oscillator $r_1 = 1$ and $r_2 = -1$ at any coupling strength is a special case of interaction. In case of weakly-coupled different oscillators when $\lambda \to 0$ the coefficients $r_{1,2} \to 0$. To demonstrate variations of $r_{1,2}$ for different oscillators compare with the case of identical oscillators we calculated $r_{1,2}$ at various parameters.

Figure 1a) exemplifies the parameters $r_1$ (curves $1, 1', 1''$) and $r_2$ (curves $2, 2', 2''$) versus the normalized coupling constant $\tilde{\lambda} = \lambda / \omega_{01}\omega_{02}\sqrt{M_1 M_2}$ in accordance with Eq.(B7). Horizontal dashed lines exemplify the case of identical oscillators. The curves $1, 2$ correspond to the case of one percent of deviation in masses and eigenfrequencies of oscillators ($M_1 = 10^{-23} g$, $M_2 = 1.01 M_1$ and $\omega_{01} = 10^{13} rad/s$, $\omega_{02} = 1.01\omega_{01}$). In its turn, the curves $1', 2'$ correspond to the case of ten percent of deviation in masses and eigenfrequencies of oscillators ($M_1 = 10^{-23} g$, $M_2 = 1.1 M_1$ and $\omega_{01} = 10^{13} rad/s$, $\omega_{02} = 1.1\omega_{01}$). Quite large deviations in parameters are characterized by the pair of curves $1'', 2''$ in this figure when $M_1 = 3\times 10^{-23} g$, $M_2 = M_1/3$ and $\omega_{01} = 0.8 \times 10^{13} rad/s$, $\omega_{02} = 10\omega_{01}$. The dissipative parameter has been chosen to be $\gamma = 0.01\omega_{01}$.

For the same parameters as those in Fig.1a) the normalized dispersions $\tilde{\sigma}_{1,2}^2(t=\infty) = \sigma_{1,2}^2(t=\infty)/\sigma_{1,2}^2(FDT)$ and covariances $\tilde{\beta}_{12}^{-1}(t=\infty) = \beta_{12}^{-1}(t=\infty)/\sqrt{\sigma_1^2(FDT)\sigma_2^2(FDT)}$ in the long-time limit of coupled oscillators interacting with different thermostats kept at different temperatures $T_1 = 300K$ and $T_2 = 700K$ versus the normalized coupling constant $\tilde{\lambda}$ are presented in

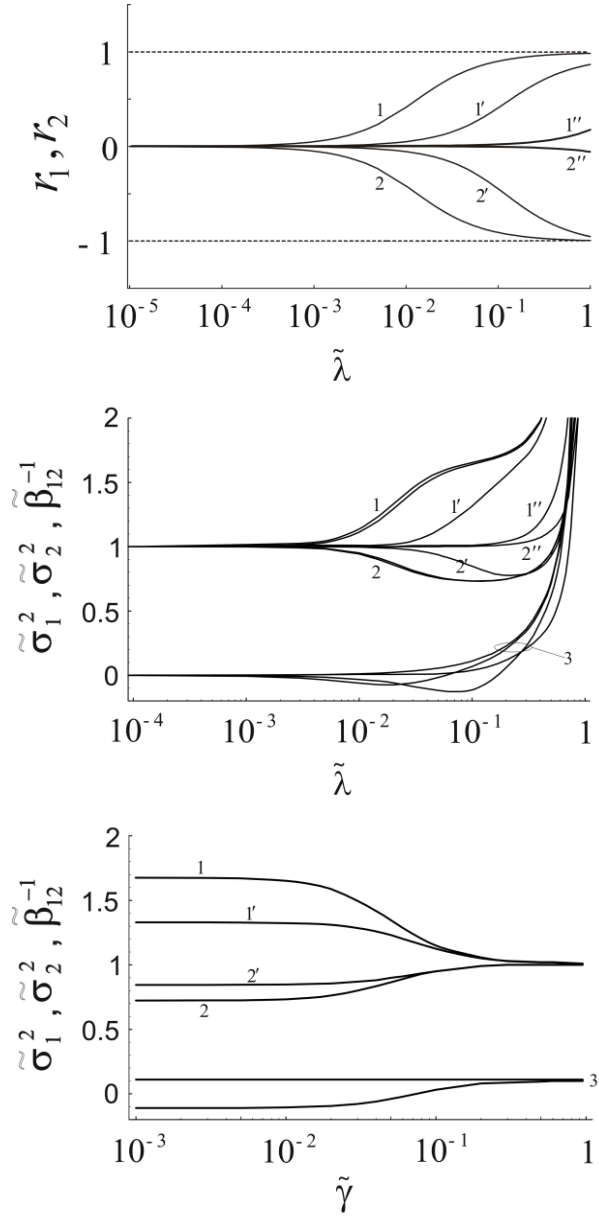

**Figure 1.** Parameters $r_1$ (curves $1,1',1''$) and $r_2$ (curves $2,2',2''$) versus $\tilde{\lambda}$ in accordance with Eq. (B7) - a). Horizontal dashed lines exemplify the case of identical coupled oscillators The curves $1,2$ for $M_2 = 1.01 M_1$, $\omega_{02} = 1.01\omega_{01}$, the curves $1',2'$ for $M_2 = 1.1 M_1$, $\omega_{02} = 1.1\omega_{01}$ at $M_1 = 10^{-23} g$ and $\omega_{01} = 10^{13} rad/s$. The pair of curves $1'',2''$ for $M_2 = M_1/3$, $\omega_{02} = 10\omega_{01}$ at $M_1 = 3\times 10^{-23} g$ and $\omega_{01} = 0.8\times 10^{13} rad/s$.

b) Normalized dispersions $\tilde{\sigma}_{1,2}^2(t=\infty)$ and covariances $\tilde{\beta}_{12}^{-1}(t=\infty)$ versus $\tilde{\lambda}$ at fixed $\gamma = 0.01\omega_{01}$ - b) and versus $\tilde{\gamma} = \gamma/\omega_{01}$ at fixed coupling constant $\tilde{\lambda} = 0.1$ - c) in accordance with Eqs.(14), (15) of coupled oscillators interacting with different thermostats kept at different temperatures $T_1 = 300K$ and $T_2 = 700K$. The curves $1,1',1''$ and $2,2',2''$ in Fig.1b) are related to the first, and second oscillators. The bunch of curves 3 correpond to covariances for all these cases. The curves $1,2$ in Fig1.c) correspond to the case of identical oscillators ($M_1 = M_2 = 10^{-23}g$, and $\omega_{01} = \omega_{02} = 10^{13} rad/s$,), the pair $1',2'$ - to the case of $M_1 = 10^{-23} g$, $M_2 = 1.1 M_1$ and $\omega_{01} = 10^{13} rad/s$, $\omega_{02} = 1.1\omega_{01}$. Two curves indicated as 3 correspond to covariances.

Fig.1b) in accordance with Eqs.(14), (15). In this figure the curves $1,1',1''$ correspond to the variance of the first oscillator, and curves $2,2',2''$ – to the second one. The bunch of curves indicated as 3 correpond to covariances for all these cases. Two pairs of nearly merging curves $1,2$ here correspond to cases of coupled identical oscillators and distinguishing up to one percent.

As well as in [47] we observe the splitting of curves 1 and 2 for different temperatures of thermostats. It is clearly seen that the distingushable splitting occures at some $\tilde{\lambda}$. But, in case of coupled oscillators of arbitrary properties we observe a more varied situation as compared with the case of interaction of identical oscillators. First of all, the larger the deviations in parameters of oscillators, the smaller is the splitting in Fig.1b). For example, for the curves $1'', 2''$ in this figure when $M_1 = 3 \times 10^{-23} g$, $M_2 = M_1/3$ and $\omega_{01} = 0.8 \times 10^{13} rad/s$, $\omega_{02} = 10\omega_{01}$ we observe $\tilde{\sigma}_{1,2}^2(t=\infty) \approx 1$ for the coupling constant up to $\tilde{\lambda} \leq 0.1$. The second feature is connected with the change in sign of the covariance $\tilde{\beta}_{12}^{-1}(t=\infty)$ with common zeroing tendency at $\tilde{\lambda} \to 0$. The principal common peculiarity in Fig.1b) is the divergent behavior of $\tilde{\sigma}_{1,2}^2(t=\infty)$ and $\tilde{\beta}_{12}^{-1}(t=\infty)$ at $\tilde{\lambda} \to 1$ as well as in case of identical oscillators [47]. To explain the observed divergent behavior of variances and covariances in general case at $\lambda = \omega_{01}\omega_{02}\sqrt{M_1 M_2}$ we consider an appropriate transformation of the Hamiltonian in Eq.(1). We note that in [40] it was shown that the two oscillator model can be effectively mapped into that of a single harmonic oscillator in a general environment plus a free harmonic oscillator. The same conclusion was done also in [43] by analysing motion equations. After changing the variables

$$z_1 = x_1\sqrt{J_1} + x_2\sqrt{J_2},$$
$$z_2 = x_1\sqrt{M_1 J_2/M_2} - x_2\sqrt{M_2 J_1/M_1}, \tag{16}$$

where $J_1 = (\omega_{01}/\omega_{01})\sqrt{M_1/M_2}$ and $J_2 = 1/J_1$, we have in case $\lambda = \omega_{01}\omega_{02}\sqrt{M_1 M_2}$ instead of Eq.(1) the following expression

$$H = P_1^2/2M + M\tilde{\omega}^2 z_1^2/2 + P_2^2/2M$$
$$+ \sum_{j=1}^{N_1}\left\{p_j^2/2m_j + m_j\omega_j^2\left[q_j - (M_2\sqrt{J_1}z_1 + M\sqrt{J_2}z_2)/(M_1 J_2 + M_2 J_1)\right]^2/2\right\} \tag{17}$$
$$+ \sum_{k=1}^{N_2}\left\{p_k^2/2m_k + m_k\omega_k^2\left[q_k - (M_1\sqrt{J_2}z_1 - M\sqrt{J_1}z_2)/(M_1 J_2 + M_2 J_1)\right]^2/2\right\},$$

where $P_1 = M\dot{z}_1$ and $P_2 = M\dot{z}_2$, $M = \sqrt{M_1 M_2}$, $\tilde{\omega} = \sqrt{\omega_{01}\omega_{02}}$.

Thus, the Hamiltonian tends to the equivalent Hamiltonian for a free effective particle and oscillator also in a general case of two oscillators of arbitrary parameters, when the coupling constant $\lambda$ tends to the value $\omega_{01}\omega_{02}\sqrt{M_1 M_2}$. That is why, due to the fictious free particle the curves are divergent in Fig.1b). It should be underlined that the new effective free particle and effective oscillator interact simulteneously with both of reservoirs, contrary to the initial system where coupled oscillators where connected to reservoirs separately. Here we would like to recall [47] that the accepted model of bilinear coupling between the selected oscillators is not perfectly appropriate to describe a strong coupling regime. Taking it into account, the divergence in Fig.1b) illustrates the restrictions imposed on the coupling between arbitrary oscillators in the frameworks of the model.

The next picture demonstrates the dependence of the steady state characteristics on the dissipation in the system. The normalized dispersions $\tilde{\sigma}_{1,2}^2(t=\infty)$ and covariances $\tilde{\beta}_{12}^{-1}(t=\infty)$ in the long-time limit of coupled oscillators interacting with different thermostats kept at different temperatures $T_1 = 300K$ and $T_2 = 700K$ versus the normalized dissipative parameter $\tilde{\gamma} = \gamma/\omega_{01}$ at fixed coupling constant $\tilde{\lambda} = 0.1$ are presented in Fig.1c). The pairs of curves $1, 1'$ and $2, 2'$ correspond to variances of the first and second oscillators. The curves $1, 2$ are related to the case of identical oscillators ($M_1 = M_2 = 10^{-23}g$, and $\omega_{01} = \omega_{02} = 10^{13} rad/s$,), the pair $1', 2'$ correspond to the case of ten percent of deviation in masses and eigenfrequencies of oscillators ($M_1 = 10^{-23}g$, $M_2 = 1.1 M_1$ and $\omega_{01} = 10^{13} rad/s$, $\omega_{02} = 1.1\omega_{01}$). Two curves indicated as $3$ correspond to covariances. In case of identical oscillators the covariance is practically a positive constant. The ten percent deviation in parameters yields in negative values of the covariance in some range of $\tilde{\gamma}$. It is seen that the larger is a dissipation parameter, the smaller is the splitting of variances.

It should be noted that the calculations for Figs.1b) and 1c) were provided for Eq.(10) at initial dispersions $\sigma_{01}^2 = \sigma_1^2(0)$ and $\sigma_{02}^2 = \sigma_2^2(0)$, where $\sigma_1^2(0) = \hbar/2M_1\omega_{01}$ and $\sigma_2^2(0) = \hbar/2M_2\omega_{02}$ mean the dispersions of isolated oscillators at $T = 0K$. In this connection we recall [52] that the fluctuation dissipation theorem provides the variances of each separate oscillator interacting with own thermostat as follows

$$\sigma_{1,2}^2(FDT) = \frac{\hbar}{\pi M_{1,2}} \int_0^\infty d\omega\, Coth\left(\frac{\hbar\omega}{2k_B T_{1,2}}\right)\frac{2\gamma\omega}{[(\omega^2 - \omega_{01,02}^2)^2 + 4\gamma^2\omega^2]}. \qquad (18)$$

It is easy to see from this expression that in case $\gamma \to 0$ and at $T_{1,2} \to 0K$ we obtain $\sigma_1^2(0) = \hbar/2M_1\omega_{01}$ and $\sigma_2^2(0) = \hbar/2M_2\omega_{02}$.

Finishing discussion of the results of our numerical study partly represented in Fig.1 we conclude that the splitting depends on difference in temperatures, parameters of oscillators, dissipation and coupling constant. Their action can be multidirectional.

Along with a study of possible deviations of steady state dispersions of coupled oscillators within a system out of equilibrium as compare with a system in equilibrium one of the main goal of this paper and of our previous works [46, 47] is connected with a relaxation dynamics of the complex system. The figure 2 illustrates the relaxation process and stationary values of variances of two coupled oscillators of different set of their parameters in contact with separate baths of different temperatures. The normalized variances $\sigma_{1,2}^2(t)/\sigma_{1,2}^2(FDT)$ of the first (the curve 1) and second (the curve 2) oscillators vesrus a time in accordance with Eqs.(14) at fixed $\tilde{\lambda}=0.1$ are shown in Fig.2) at different temperatures of baths $T_1=300K$ and $T_2=900K$. Normalization was done to the variances $\sigma_{1,2}^2(FDT)$ of oscillators in equilibrium with their baths at $T_1$ and $T_2$ in accordance with the fluctuation dissipation theorem.

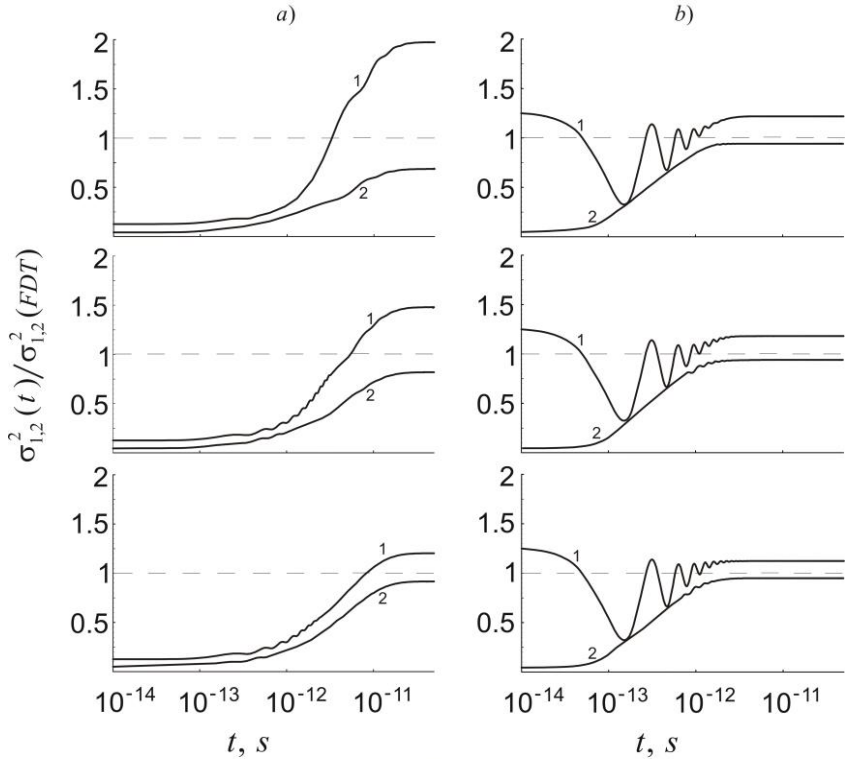

Fig.2

**Figure 2.** Temporal dynamics of normalized variances $\sigma_{1,2}^2(t)/\sigma_{1,2}^2(FDT)$ of the first (the curve 1) and second (the curve 2) oscillators in accordance with Eqs.(14) at fixed $\tilde{\lambda}=0.1$ and different temperatures of baths $T_1=300K$ and $T_2=900K$. Normalization was done to the variances $\sigma_{1,2}^2(FDT)$ of oscillators in equilibrium with their baths at $T_1$ and $T_2$ in accordance with the FDT. The graphics were prepared at $M_2=M_1$, $\omega_{02}=\omega_{01}$ for the upper pictures, at $M_2=1.1M_1$, $\omega_{02}=1.1\omega_{01}$ for the middle pictures, and at $M_2=1.2M_1$, $\omega_{02}=1.2\omega_{01}$ for the low pictures. Initial dispersions $\sigma_{01}^2=\sigma_1^2(0)$, $\sigma_{02}^2=\sigma_2^2(0)$ and $\gamma_1=\gamma_2=\gamma=0.01\omega_{01}$ - a). Initial dispersions $\sigma_{01}^2=10\sigma_1^2(0)$, $\sigma_{02}^2=\sigma_2^2(0)$ and $\gamma_1=\gamma_2=\gamma=0.1\omega_{01}$ - b).

The pictures in figures 2a) and 2b) were prepared at $M_2 = M_1$, $\omega_{02} = \omega_{01}$ for the upper pictures, at $M_2 = 1.1M_1$, $\omega_{02} = 1.1\omega_{01}$ for the middle pictures, and at $M_2 = 1.2M_1$, $\omega_{02} = 1.2\omega_{01}$ for the low pictures. In other words, the figures correspond to the ten and twenty percent deviation in parameters of oscillators comparing with the case of identical oscillators. The difference between Fig.2a) and 2b) is in different initial dispersions and different dissipative parameters. The case of initially cold oscillators with dispersions $\sigma_{01}^2 = \sigma_1^2(0)$ and $\sigma_{02}^2 = \sigma_2^2(0)$ for Eq.(10) is illustrated by Fig.2a) at $\gamma_1 = \gamma_2 = \gamma = 0.01\omega_{01}$. Other case of initial dispersions $\sigma_{01}^2 = 10\sigma_1^2(0)$ and $\sigma_{02}^2 = \sigma_2^2(0)$ is illustrated by Fig.2b) at $\gamma_1 = \gamma_2 = \gamma = 0.1\omega_{01}$. For numerical calulations we have chosen the parameters of the first oscillators to be $M_1 = 10^{-23} g$ and $\omega_{01} = 10^{13} rad/s$, relating to the typical characteristics for solids. It is seen from Fig.2 (and from Fig.1, too) that the larger the difference in parameters between the selected oscillators, the smaller the deviations of dispersions in this case from dispersions of oscillators in total equilibrium. From our point of view the physical sense is as follows: the most effective channels of interactions of subsystems within a complex system can be realized mainly at most fitted degrees of freedom (eigenmodes). This situation is similar to the resonance increase in dispersion interactions between two bodies under the additional action of a third body or background at coinciding eigenfrequencies within a system out of total equilibrium [7, 11].

So far, we consider the temperature range $k_B T \geq \hbar \omega_{01,02}$, where effects relating to the difference in temperature are of major interest. In the low-temperature regime $k_B T \leq \hbar \omega_{01,02}$ or even at $T = 0K$ the relaxation dynamics of the system under our study is also not trivial. Figure 3 exemplifies the relaxation process of coupled oscillators interacting with baths of equal temperatures - a), and interacting with baths at different temperatures - b). In this figure the temporal behaviour of normalized variances $\sigma_{1,2}^2(t)/\sigma_{1,2}^2(FDT)$ of the first (the curve 1) and second (the curve 2) oscillators in accordance with Eqs.(14) are shown at fixed coupling constant $\tilde{\lambda} = 0.2$ and at equal temperatures of baths $T_1 = T_2 = 0K$ in Fig.3a), and at different temperatures $T_1 = 0K$ and $T_2 = 100K$ in Fig.3b). Normalization was done to the variances $\sigma_{1,2}^2(FDT)$ of oscillators in equilibrium separately with their baths at $T_1$ and $T_2$ due to the FDT. The pictures in figures 3a) and 3b) were prepared at $M_2 = M_1$, $\omega_{02} = \omega_{01}$ for the upper pictures, at $M_2 = 1.2M_1$, $\omega_{02} = 1.2\omega_{01}$ for the middle pictures, and at $M_2 = 1.5M_1$, $\omega_{02} = 1.5\omega_{01}$ for the low pictures. For all the cases initial dispersions are to be selected so that $\sigma_{01}^2 = \sigma_1^2(0)$ and $\sigma_{02}^2 = 2\sigma_2^2(0)$ at the common dissipative parameter $\gamma = 0.05\omega_{01}$. For calculations the parameters $M_1 = 10^{-23} g$ and $\omega_{01} = 10^{13} rad/s$ have been chosen as for previous examples.

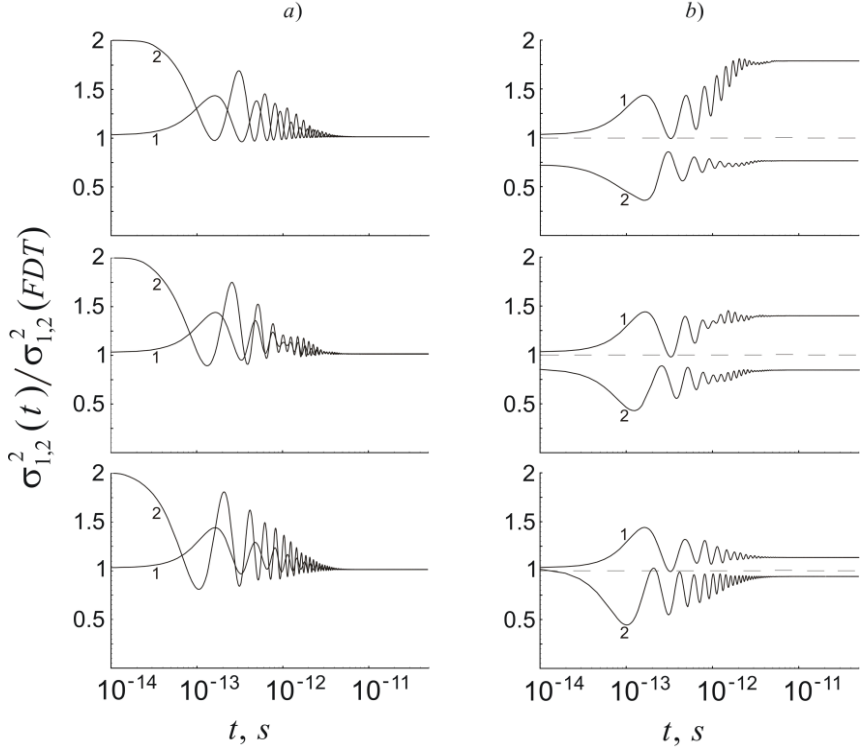

Fig.3

**Figure 3.** Temporal behaviour of normalized variances $\sigma_{1,2}^2(t)/\sigma_{1,2}^2(FDT)$ of the first (the curve 1) and second (the curve 2) oscillators at fixed coupling constant $\tilde{\lambda} = 0.2$ in accordance with Eqs.(14) at equal temperatures of baths $T_1 = T_2 = 0K$ - a), and at different temperatures $T_1 = 0K$ and $T_2 = 100K$ - b). Normalization was done to the variances $\sigma_{1,2}^2(FDT)$ of oscillators in equilibrium with their baths at $T_1$ and $T_2$ in accordance with the FDT. The pictures in figures 3a) and 3b) were prepared at $M_2 = M_1$, $\omega_{02} = \omega_{01}$ for the upper pictures, at $M_2 = 1.2M_1$, $\omega_{02} = 1.2\omega_{01}$ for the middle pictures, and at $M_2 = 1.5M_1$, $\omega_{02} = 1.5\omega_{01}$ for the low pictures. For all the cases initial dispersions are to be selected so that $\sigma_{01}^2 = \sigma_1^2(0)$ and $\sigma_{02}^2 = 2\sigma_2^2(0)$ at the common dissipative parameter $\gamma = 0.05\omega_{01}$.

It follows clearly from Fig.3a) that in a system in equilibrium at $T_{1,2} = 0K$ the variances are in accordance with the fluctuation dissipation theorem as it must be. It is seen from Figs.3a) and 3b) that the difference in temperatures in the low-temperature regime yields in the deviations of dispersions from their values corresponding to the case of total equilibrium. We also observe that the larger is the difference in parameters of coupled oscillators, the smaller is the difference of the steady state dispersions from their equilibrium values.

Thereby, our studies show that in frameworks of the approach of factorized initial conditions, bilinear coupling and Ohmic dissipation the system of two coupled arbitrary oscillators relaxes to the steady state independently on given initial dispersions and difference in temperatures. In case of different temperatures of thermostats the stationary values of variances and covariance may be different from values corresponding to the case of total equilibrium. These deviations depend on the coupling between the selected oscillators and their parameters, on the dissipation factor and difference in temperatures.

## IV. Conclusion.

Our paper addresses the problem of the relaxation dynamics of complex open quantum systems to quasi-equilibrium states. Using the path integral methods we found an analytical expression for time-dependent density matrix of two arbitrary coupled quantum oscillators of arbitrary parameters interacting with separate reservoirs of oscillators. As for initial conditions we assumed a factorized form for the initial density matrix. Temporal dynamics of spatial variances and covariances of the oscillators from any given time up to quasi-equilibrium steady states is studied. It is demonstrated that asymptotic spatial variances of two arbitrary oscillators and their covariances achieve stationary values in the long-time limit. A comparison of steady state characteristics of coupled oscillators with those predicted by the fluctuation dissipation theorem (FDT) is performed. From the general formula we can to obtain all derived before expressions for the weak-coupling approach and for a system of two identical oscillators [46, 47]. Time-dependent spatial variances and covariance were investigated analytically and numerically. In the short-time limit the variances are identical to their initially given values, as it must be. It was confirmed in general case that in the weak coupling approach the variances in the long-time limit are always asymptotically in accordance at some accuracy with the fluctuation dissipation theorem despite on their initial values. We numerically investigated the high- and low temperature limits of interactions within the system under our study. It was shown for general case of arbitrary oscillators the stationary values of variances differ from the case of total equilibrium due to effective coupling of oscillators simultaneously with thermal reservoirs of different temperatures. It was shown for general case the larger the difference in temperatures of thermal baths, the larger is the difference of the steady state variances of coupled oscillators comparing with values given by the fluctuation dissipation theorem. Besides, it was shown that the larger the difference in masses and eigenfrequencies of coupled oscillators, the smaller are the deviations of stationary variances from those given by the FDT at fixed coupling strength and fixed difference in temperatures between thermal baths. In framework of the accepted model of the bilinear coupling, Ohmic dissipation and at any oscillators' parameters the variances and covariances have divergent character at strong couplings. A related explanation is suggested for general case based on transformation of the initial Hamiltonian to the Hamiltonian containing the term for a free particle. We demonstrated that the dissipative parameter plays a crucial role both for a total time of relaxation, and for deviations of dispersions of coupled oscillators within a system out of equilibrium compare with an equilibrium case. Thus, in our papers we considered the problem on relaxation dynamics of two coupled oscillators interacting with different thermostats in the weak coupling limit [46], in a special case of identical oscillators [47], and in a general case of arbitrary oscillators and arbitrary dissipation in this paper. Besides, the steady states of two coupled oscillators, their energies and the energy of interaction were studied in [45].

## Appendix A. Time-dependent functions in Eq.(9).

The classical action in Eq.(9) in new variables is expressed as follows

$$\tilde{S}_{cl}^{(1)}(t) + \tilde{S}_{cl}^{(2)}(t) + \tilde{S}_{cl}^{(12)}(t) = \int_0^t dt' \{ M_1 \dot{\tilde{X}}_1 \dot{\tilde{\xi}}_1 / 2 - M_1 \omega_1^2 \tilde{X}_1 \tilde{\xi}_1 / 2 - M_1 \gamma_1 \dot{\tilde{X}}_1 \tilde{\xi}_1 \\ + M_2 \dot{\tilde{X}}_2 \dot{\tilde{\xi}}_2 / 2 - M_2 \omega_2^2 \tilde{X}_2 \tilde{\xi}_2 / 2 - M_2 \gamma_2 \dot{\tilde{X}}_2 \tilde{\xi}_2 + (\lambda / 2)[\tilde{X}_1(t')\tilde{\xi}_2(t') + \tilde{X}_2(t')\tilde{\xi}_1(t')] \}, \quad \text{(A1)}$$

where $\tilde{X}_{1,2}$ and $\tilde{\xi}_{1,2}$ are the classical paths given by Eqs.(A6), (A8)-(A11) from [46].

After an integration in Eq.(A1) we have

$$\tilde{S}_{cl}^{(1)} + \tilde{S}_{cl}^{(2)} = D_1(X_{f1}\xi_{f1}) + [D_5 + D_5'](X_{f2}\xi_{f1}) + [D_6 + D_6'](X_{f1}\xi_{f2}) + D_1'(X_{f2}\xi_{f2}) + \\ + D_2(X_{i1}\xi_{f1}) + D_3(X_{f1}\xi_{i1}) + D_4(X_{i1}\xi_{i1}) + D_2'(X_{i2}\xi_{f2}) + D_3'(X_{f2}\xi_{i2}) + D_4'(X_{i2}\xi_{i2}) + \\ + [D_7 + D_7'](X_{i1}\xi_{f2}) + [D_8 + D_8'](X_{i2}\xi_{f1}) + [D_9 + D_9'](X_{f1}\xi_{i2}) + [D_{10} + D_{10}'](X_{f2}\xi_{i1}) + \\ + [D_{11} + D_{11}'](X_{i1}\xi_{i2}) + [D_{12} + D_{12}'](X_{i2}\xi_{i1}) \quad \text{(A2)}$$

$$\tilde{S}_{cl}^{(12)} = \Pi_1(X_{f1}\xi_{f1}) + \Pi_2(X_{f1}\xi_{f2}) + \Pi_3(X_{f2}\xi_{f1}) + \Pi_4(X_{f2}\xi_{f2}) + \\ + \Pi_5(X_{f1}\xi_{i1}) + \Pi_6(X_{f1}\xi_{i2}) + \Pi_7(X_{f2}\xi_{i1}) + \Pi_8(X_{f2}\xi_{i2}) + \\ + \Pi_9(X_{i1}\xi_{f1}) + \Pi_{10}(X_{i1}\xi_{f2}) + \Pi_{11}(X_{i2}\xi_{f1}) + \Pi_{12}(X_{i2}\xi_{f2}) + \\ + \Pi_{13}(X_{i1}\xi_{i1}) + \Pi_{14}(X_{i1}\xi_{i2}) + \Pi_{15}(X_{i2}\xi_{i1}) + \Pi_{16}(X_{i2}\xi_{i2}) \quad \text{(A3)}$$

where the time-dependent functions $D_k \equiv D_k(t), D_k' \equiv D_k'(t), \ (k = 1,...,12)$ are as follows

$$D_1(t) = (M_1/2)[n_1\bar{n}_1 b_1 - r_1 n_1 \bar{n}_2 b_9 - r_1 n_2 \bar{n}_1 b_5 + r_1^2 n_2 \bar{n}_2 b_{13}]$$
$$+ (M_2/2)[n_1\bar{n}_1 b'_{13} - r_1 n_1 \bar{n}_2 b'_9 - r n_2 \bar{n}_1 b'_5 + r_1^2 n_2 \bar{n}_2 b'_1] \; ,$$
$$D'_1(t) = (M_1/2)[r_2^2 n_1\bar{n}_1 b_1 - r_2 n_1 \bar{n}_2 b_9 - r_2 n_2 \bar{n}_1 b_5 + n_2 \bar{n}_2 b_{13}]$$
$$+ (M_2/2)[r_2^2 n_1\bar{n}_1 b'_{13} - r_2 n_1 \bar{n}_2 b'_9 - r_2 n_2 \bar{n}_1 b'_5 + n_2 \bar{n}_2 b'_1] \; ,$$
(A4)

$$D_2(t) = (M_1/2)[-m_1\bar{n}_1 b_1 + r_1 m_1 \bar{n}_2 b_9 + \bar{n}_1 \ell b_2 - r_1 \bar{n}_2 \ell b_{10} + r_1 m_2 \bar{n}_1 b_5 - r_1^2 m_2 \bar{n}_2 b_{13} - r_1 \bar{n}_1 \ell b_6 + r_1^2 \bar{n}_2 \ell b_{14}]$$
$$+ (M_2/2)[-m_1\bar{n}_1 b'_{13} + r_1 m_1 \bar{n}_2 b'_9 + \bar{n}_1 \ell b'_{14} - r_1 \bar{n}_2 \ell b'_{10} + r_1 m_2 \bar{n}_1 b'_5 - r_1^2 m_2 \bar{n}_2 b'_{13} - r_1 \bar{n}_1 \ell b'_6 + r_1^2 \bar{n}_2 \ell b'_2] \; ,$$
$$D'_2(t) = (M_1/2)[-r_2^2 m_1\bar{n}_1 b_1 + r_2 m_1 \bar{n}_2 b_9 + r_2^2 \bar{n}_1 \ell b_2 - r_2 \bar{n}_2 \ell b_{10} + r_2 m_2 \bar{n}_1 b_5 - m_2 \bar{n}_2 b_{13} - r_2 \bar{n}_1 \ell b_6 + \bar{n}_2 \ell b_{14}]$$
$$+ (M_2/2)[-r_2^2 m_1\bar{n}_1 b'_{13} + r_2 m_1 \bar{n}_2 b'_9 + r_2^2 \bar{n}_1 \ell b'_{14} - r_2 \bar{n}_2 \ell b'_{10} + r_2 m_2 \bar{n}_1 b'_5 - m_2 \bar{n}_2 b'_1 - r_2 \bar{n}_1 \ell b'_6 + \bar{n}_2 \ell b'_2] \; ,$$
(A5)

$$D_3(t) = (M_1/2)[-m_1 n_1 b_1 + n_1 \ell b_3 + r_1 m_2 n_1 b_9 - r_1 n_1 \ell b_{11} + r_1 m_1 n_2 b_5 - r_1 n_2 \ell b_7 - r_1^2 m_2 n_2 b_{13} + r_1^2 n_2 \ell b_{15}]$$
$$+ (M_2/2)[-m_1 n_1 b'_{13} + n_1 \ell b'_{15} + r_1 m_2 n_1 b'_9 - r_1 n_1 \ell b'_{11} + r_1 m_1 n_2 b'_5 - r_1 n_2 \ell b'_8 - r_1^2 m_2 n_2 b'_1 + r_1^2 n_2 \ell b'_3] \; ,$$
$$D'_3(t) = (M_1/2)[-r_2^2 m_1 n_1 b_1 + r_2^2 n_1 \ell b_3 + r_2 m_2 n_1 b_9 - r_2 n_1 \ell b_{11} + r_2 m_1 n_2 b_5 - r_2 n_2 \ell b_7 - m_2 n_2 b_{13} + n_2 \ell b_{15}]$$
$$+ (M_2/2)[-r_2^2 m_1 n_1 b'_{13} + r_2^2 n_1 \ell b'_{15} + r_2 m_2 n_1 b'_9 - r_2 n_1 \ell b'_{11} + r_2 m_1 n_2 b'_5 - r_2 n_2 \ell b'_8 - m_2 n_2 b'_1 + n_2 \ell b'_3],$$
(A6)

$$D_4(t) = (M_1/2)[m_1^2 b_1 - m_1 \ell b_3 - r_1 m_1 m_2 b_9 + r_1 m_1 \ell b_{11} - m_1 \ell b_2 + \ell^2 b_4 + r_1 m_2 \ell b_{10} - r_1^2 \ell^2 b_{12}$$
$$- r_1 m_1 m_2 b_5 + r_1 m_2 \ell b_7 + r_1^2 m_2^2 b_{13} - r_1^2 m_2 \ell b_{15} + r_1 m_1 \ell b_6 - r_1^2 \ell^2 b_8 - r_1^2 m_2 \ell b_{14} + r_1^2 \ell^2 b_{16}]$$
$$+ (M_2/2)[m_1^2 b'_{13} - m_1 \ell b'_{15} - r_1 m_1 m_2 b'_9 + r_1 m_1 \ell b'_{11} - m_1 \ell b'_{14} + \ell^2 b'_{16} + r_1 m_2 \ell b'_{10} - r_1^2 \ell^2 b'_{12}$$
$$- r_1 m_1 m_2 b'_5 + r_1 m_2 \ell b'_8 + r_1^2 m_2^2 b'_1 - r_1^2 m_2 \ell b'_3 + r_1 m_1 \ell b'_6 - r_1^2 \ell^2 b'_7 - r_1^2 m_2 \ell b'_2 + r_1^2 \ell^2 b'_4],$$
(A7)

$$D'_4(t) = (M_1/2)[r_2^2 m_1^2 b_1 - r_2^2 m_1 \ell b_3 - r_2 m_1 m_2 b_9 + r_2 m_1 \ell b_{11} - r_2^2 m_1 \ell b_2 + r_2^2 \ell^2 b_4 + r_2 m_2 \ell b_{10}$$
$$- r_2 \ell^2 b_{12} - r_2 m_1 m_2 b_5 + r_2 m_2 \ell b_7 + m_2^2 b_{13} - m_2 \ell b_{15} + r_2 m_1 \ell b_6 - r_2 \ell^2 b_8 - m_2 \ell b_{14} + \ell^2 b_{16}]$$
$$+ (M_2/2)[r_2^2 m_1^2 b'_{13} - r_2^2 m_1 \ell b'_{15} - r_2 m_1 m_2 b'_9 + r_2 m_1 \ell b'_{11} - r_2^2 m_1 \ell b'_{14} + r_2^2 \ell^2 b'_{16} + r_2 m_2 \ell b'_{10}$$
$$- r_2 \ell^2 b'_{12} - r_2 m_1 m_2 b'_5 + r_2 m_2 \ell b'_8 + m_2^2 b'_1 - m_2 \ell b'_3 + r_2 m_1 \ell b'_6 - r_2 \ell^2 b'_7 - m_2 \ell b'_2 + \ell^2 b'_4],$$
(A8)

$$D_5(t) + D'_5(t) = (M_1/2)[-r_2 n_1 \bar{n}_1 b_1 + r_1 r_2 n_1 \bar{n}_2 b_9 + n_2 \bar{n}_1 b_5 - r_1 n_2 \bar{n}_2 b_{13}]$$
$$+ (M_2/2)[-r_2 n_1 \bar{n}_1 b'_{13} + r_1 r_2 n_1 \bar{n}_2 b'_9 + n_2 \bar{n}_1 b'_5 - r_1 n_2 \bar{n}_2 b'_1],$$
$$D_6(t) + D'_6(t) = (M_1/2)[-r_2 n_1 \bar{n}_1 b_1 + n_1 \bar{n}_2 b_9 + r_1 r_2 n_2 \bar{n}_1 b_5 - r_1 n_2 \bar{n}_2 b_{13}]$$
$$+ (M_2/2)[-r_2 n_1 \bar{n}_1 b'_{13} + n_1 \bar{n}_2 b'_9 + r_1 r_2 n_2 \bar{n}_1 b'_5 - r_1 n_2 \bar{n}_2 b'_1],$$
(A9)

$$D_7(t) + D_7'(t) = (M_1/2)[r_2 m_1 \bar{n}_1 b_1 - m_1 \bar{n}_2 b_9 - r_2 \bar{n}_1 \ell b_2 + \bar{n}_2 \ell b_{10} - r_1 r_2 m_2 \bar{n}_1 b_5$$
$$+ r_1 m_2 \bar{n}_2 b_{13} + r_1 r_2 \bar{n}_1 \ell b_6 - r_1 \bar{n}_2 \ell b_{14}]$$
$$+ (M_2/2)[r_2 m_1 \bar{n}_1 b_{13}' - m_1 \bar{n}_2 b_9' - r_2 \bar{n}_1 \ell b_{14}' + \bar{n}_2 \ell b_{10}' - r_1 r_2 m_2 \bar{n}_1 b_5'$$
$$+ r_1 m_2 \bar{n}_2 b_1' + r_1 r_2 \bar{n}_1 \ell b_6' - r_1 \bar{n}_2 \ell b_2'],$$

$$D_8(t) + D_8'(t) = (M_1/2)[r_2 m_1 \bar{n}_1 b_1 - r_1 r_2 m_1 \bar{n}_2 b_9 - r_2 \bar{n}_1 \ell b_2 + r_1 r_2 \bar{n}_2 \ell b_{10}$$
$$- m_2 \bar{n}_1 b_5 + r_1 m_2 \bar{n}_2 b_{13} + \bar{n}_1 \ell b_6 - r_1 \bar{n}_2 \ell b_{14}]$$
$$+ (M_2/2)[r_2 m_1 \bar{n}_1 b_{13}' - r_1 r_2 m_1 \bar{n}_2 b_9' - r_2 \bar{n}_1 \ell b_{14}' + r_1 r_2 \bar{n}_2 \ell b_{10}'$$
$$- m_2 \bar{n}_1 b_5' + r_1 m_2 \bar{n}_2 b_1' + \bar{n}_1 \ell b_6' - r_1 \bar{n}_2 \ell b_2'],$$
(A10)

$$D_9(t) + D_9'(t) = (M_1/2)[r_2 m_1 n_1 b_1 - r_2 n_1 \ell b_3 - m_2 n_1 b_9 + n_1 \ell b_{11} - r_1 r_2 m_1 n_2 b_5$$
$$+ r_1 r_2 n_2 \ell b_7 + r_1 m_2 n_2 b_{13} - r_1 n_2 \ell b_{15}]$$
$$(M_2/2)[r_2 m_1 n_1 b_{13}' - r_2 n_1 \ell b_{15}' - m_2 n_1 b_9' + n_1 \ell b_{11}' - r_1 r_2 m_1 n_2 b_5'$$
$$+ r_1 r_2 n_2 \ell b_8' + r_1 m_2 n_2 b_1' - r_1 n_2 \ell b_3'],$$

$$D_{10}(t) + D_{10}'(t) = (M_1/2)[r_2 m_1 n_1 b_1 - r_2 n_1 \ell b_3 - r_1 r_2 m_2 n_1 b_9 + r_1 r_2 n_1 \ell b_{11}$$
$$- m_1 n_2 b_5 + n_2 \ell b_7 + r_1 m_2 n_2 b_{13} - r_1 n_2 \ell b_{15}]$$
$$+ (M_2/2)[r_2 m_1 n_1 b_{13}' - r_2 n_1 \ell b_{15}' - r_1 r_2 m_2 n_1 b_9' + r_1 r_2 n_1 \ell b_{11}'$$
$$- m_1 n_2 b_5' + n_2 \ell b_8' + r_1 m_2 n_2 b_1' - r_1 n_2 \ell b_3'],$$
(A11)

$$D_{11}(t) + D_{11}'(t) = (M_1/2)[-r_2 m_1^2 b_1 + r_2 m_1 \ell b_3 + m_2 m_1 b_9 - m_1 \ell b_{11} + r_2 m_1 b_2 - r_2 \ell^2 b_4 - m_2 \ell b_{10}$$
$$+ \ell^2 b_{12} + r_1 r_2 m_1 m_2 b_5 - r_1 r_2 m_2 \ell b_7 - r_1 m_2^2 b_{13} + r_1 m_2 \ell b_{15} - r_1 r_2 m_1 \ell b_6 + r_1 r_2 \ell^2 b_8 + r_1 m_2 \ell b_{14} - r_1 \ell^2 b_{16}],$$
$$+ (M_2/2)[-r_2 m_1^2 b_{13}' + r_2 m_1 \ell b_{15}' + m_2 m_1 b_9' - m_1 \ell b_{11}' + r_2 m_1 \ell b_{14}' - r_2 \ell^2 b_{16}' - m_2 \ell b_{10}'$$
$$+ \ell^2 b_{12}' + r_1 r_2 m_1 m_2 b_5' - r_1 r_2 m_2 \ell b_8' - r_1 m_2^2 b_1' + r_1 m_2 \ell b_3' - r_1 r_2 m_1 \ell b_6' + r_1 r_2 \ell^2 b_7' + r_1 m_2 \ell b_2' - r_1 \ell^2 b_4'],$$
(A12)

$$D_{12}(t) + D_{12}'(t) = (M_1/2)[-r_2 m_1^2 b_1 + r_2 m_1 \ell b_3 + r_1 r_2 m_2 m_1 b_9 - r_1 r_2 m_1 \ell b_{11} + r_2 m_1 b_2 - r_2 \ell^2 b_4$$
$$- r_1 r_2 m_2 \ell b_{10} + r_1 r_2 \ell^2 b_{12} + m_1 m_2 b_5 - m_2 \ell b_7 - r_1 m_2^2 b_{13} + r_1 m_2 \ell b_{15} - m_1 \ell b_6 + \ell^2 b_8 + r_1 m_2 \ell b_{14} - r_1 \ell^2 b_{16}]$$
$$+ (M_1/2)[-r_2 m_1^2 b_{13}' + r_2 m_1 \ell b_{15}' + r_1 r_2 m_2 m_1 b_9' - r_1 r_2 m_1 \ell b_{11}' + r_2 m_1 \ell b_{14}' - r_2 \ell^2 b_{16}'$$
$$- r_1 r_2 m_2 \ell b_{10}' + r_1 r_2 \ell^2 b_{12}' + m_1 m_2 b_5' - m_2 \ell b_8' - r_1 m_2^2 b_1' + r_1 m_2 \ell b_3' - m_1 \ell b_6' + \ell^2 b_7' + r_1 m_2 \ell b_2' - r_1 \ell^2 b_4'],$$
(A13)

where

$$n_{1,2} \equiv n_{1,2}(t) = \exp(\gamma t)/[(1-r_1 r_2)\sin(\Omega_{1,2} t)], \quad \bar{n}_{1,2} \equiv \bar{n}_{1,2}(t) = \exp(-\gamma t)/[(1-r_1 r_2)\sin(\Omega_{1,2} t)],$$
$$m_{1,2} \equiv m_{1,2}(t) = \cot(\Omega_{1,2} t)/(1-r_1 r_2), \quad \ell = 1/(1-r_1 r_2), \quad b_k \equiv b_k(t), \quad b_k' \equiv b_k'(t), \quad (k=1,\ldots,16)$$
(A14)

$$b_1 = \Omega_1^2 s_1 - \tilde{\omega}_{01}^2 s_2 - 2\Omega_1\gamma s_5, \ b_2 = -\Omega_1^2 s_5 - \tilde{\omega}_{01}^2 s_5 - \Omega_1\gamma(s_1 - s_2),$$
$$b_3 = -\Omega_1^2 s_5 - \tilde{\omega}_{01}^2 s_5 - \Omega_1\gamma(s_1 - s_2), \ b_4 = \Omega_1^2 s_2 - \tilde{\omega}_{01}^2 s_1 + 2\Omega_1\gamma s_5,$$
$$b_5 = r_2[\Omega_1\Omega_2 s_7 - \tilde{\omega}_{01}^2 s_{10} - \Omega_1\gamma s_8 - \Omega_2\gamma s_9], \ b_6 = r_2[-\Omega_1\Omega_2 s_8 - \tilde{\omega}_{01}^2 s_9 - \Omega_1\gamma s_7 + \Omega_2\gamma s_{10}],$$
$$b_7 = r_2[-\Omega_1\Omega_2 s_9 - \tilde{\omega}_{01}^2 s_8 + \Omega_1\gamma s_{10} - \Omega_2\gamma s_7], \ b_8 = r_2[\Omega_1\Omega_2 s_{10} - \tilde{\omega}_{01}^2 s_7 + \Omega_1\gamma s_9 + \Omega_2\gamma s_8],$$
$$b_9 = r_2[\Omega_1\Omega_2 s_{11} - \tilde{\omega}_{01}^2 s_{14} - \Omega_2\gamma s_{12} - \Omega_1\gamma s_{13}], \ b_{10} = r_2[-\Omega_1\Omega_2 s_{12} - \tilde{\omega}_{01}^2 s_{13} - \Omega_2\gamma s_{11} + \Omega_1\gamma s_{14}], \quad \text{(A15)}$$
$$b_{11} = r_2[-\Omega_1\Omega_2 s_{13} - \tilde{\omega}_{01}^2 s_{12} + \Omega_2\gamma s_{14} - \Omega_1\gamma s_{11}], \ b_{12} = r_2[\Omega_1\Omega_2 s_{14} - \tilde{\omega}_{01}^2 s_{11} + \Omega_2\gamma s_{13} + \Omega_1\gamma s_{12}],$$
$$b_{13} = r_2^2[\Omega_2^2 s_3 - \tilde{\omega}_{01}^2 s_4 - 2\Omega_2\gamma s_6], \ b_{14} = r_2^2[-\Omega_2^2 s_6 - \tilde{\omega}_{01}^2 s_6 + \Omega_2\gamma(s_4 - s_3)],$$
$$b_{15} = r_2^2[-\Omega_2^2 s_6 - \tilde{\omega}_{01}^2 s_6 + \Omega_2\gamma(s_4 - s_3)], \ b_{16} = r_2^2[\Omega_2^2 s_4 - \tilde{\omega}_{01}^2 s_3 + 2\Omega_2\gamma s_6],$$

$$b_1' = \Omega_2^2 s_3 - \tilde{\omega}_{02}^2 s_4 - 2\Omega_2\gamma s_6, \ b_2' = -\Omega_2^2 s_6 - \tilde{\omega}_{02}^2 s_6 - \Omega_2\gamma(s_3 - s_4),$$
$$b_3' = -\Omega_2^2 s_6 - \tilde{\omega}_{02}^2 s_6 - \Omega_2\gamma(s_3 - s_4), \ b_4' = \Omega_2^2 s_4 - \tilde{\omega}_{02}^2 s_3 + 2\Omega_2\gamma s_6,$$
$$b_5' = r_1[\Omega_1\Omega_2 s_7 - \tilde{\omega}_{02}^2 s_{10} - \Omega_1\gamma s_8 - \Omega_2\gamma s_9], \ b_6' = r_1[-\Omega_1\Omega_2 s_8 - \tilde{\omega}_{02}^2 s_9 - \Omega_1\gamma s_7 + \Omega_2\gamma s_{10}],$$
$$b_7' = r_1[\Omega_1\Omega_2 s_{10} - \tilde{\omega}_{02}^2 s_7 + \Omega_1\gamma s_9 + \Omega_2\gamma s_8], \ b_8' = r_1[-\Omega_1\Omega_2 s_9 - \tilde{\omega}_{02}^2 s_8 + \Omega_1\gamma s_{10} - \Omega_2\gamma s_7),$$
$$b_9' = r_1[\Omega_1\Omega_2 s_{11} - \tilde{\omega}_{02}^2 s_{14} - \Omega_2\gamma s_{12} - \Omega_1\gamma s_{13}], \ b_{10}' = r_1[-\Omega_1\Omega_2 s_{12} - \tilde{\omega}_{02}^2 s_{13} - \Omega_2\gamma s_{11} + \Omega_1\gamma s_{14}], \quad \text{(A16)}$$
$$b_{11}' = r_1[-\Omega_1\Omega_2 s_{13} - \tilde{\omega}_{02}^2 s_{12} + \Omega_2\gamma s_{14} - \Omega_1\gamma s_{11}], \ b_{12}' = r_1[\Omega_1\Omega_2 s_{14} - \tilde{\omega}_{02}^2 s_{11} + \Omega_2\gamma s_{13} + \Omega_1\gamma s_{12}],$$
$$b_{13}' = r_1^2[\Omega_1^2 s_1 - \tilde{\omega}_{02}^2 s_2 - 2\Omega_1\gamma s_5], \ b_{14}' = r_1^2[-\Omega_1^2 s_5 - \tilde{\omega}_{02}^2 s_5 + \Omega_1\gamma(s_2 - s_1)],$$
$$b_{15}' = r_1^2[-\Omega_1^2 s_5 - \tilde{\omega}_{02}^2 s_5 + \Omega_1\gamma(s_2 - s_1)], \ b_{16}' = r_1^2[\Omega_1^2 s_2 - \tilde{\omega}_{02}^2 s_1 + 2\Omega_1\gamma s_5],$$

where $\tilde{\omega}_{01,02}^2 = (\omega_{01,02}^2 - \gamma^2)$ for brevity.

For the time-dependent functions $\Pi_k \equiv \Pi_k(t)$, $(k = 1,...,16)$ in Eq.(A3) we have

$$\Pi_1(t) = \lambda[r_1 n_1 \bar{n}_1 s_2 - r_1(1 + r_1 r_2) n_1 \bar{n}_2 s_{14}/2 - r_1(1 + r_1 r_2) n_2 \bar{n}_1 s_{10}/2 + r_2 r_1^2 n_2 \bar{n}_2 s_4],$$
$$\Pi_2(t) = \lambda[-r_1 r_2 n_1 \bar{n}_1 s_2 + (1 + r_1 r_2) n_1 \bar{n}_2 s_{14}/2 + r_1 r_2(1 + r_1 r_2) n_2 \bar{n}_1 s_{10}/2 - r_1 r_2 n_2 \bar{n}_2 s_4],$$
$$\Pi_3(t) = \lambda[-r_1 r_2 n_1 \bar{n}_1 s_2 + r_1 r_2(1 + r_1 r_2) n_1 \bar{n}_2 s_{14}/2 + (1 + r_1 r_2) n_2 \bar{n}_1 s_{10}/2 - r_1 r_2 n_2 \bar{n}_2 s_4], \quad \text{(A17)}$$
$$\Pi_4(t) = \lambda[r_1 r_2^2 n_1 \bar{n}_1 s_2 - r_2(1 + r_1 r_2) n_1 \bar{n}_2 s_{14}/2 - r_2(1 + r_1 r_2) n_2 \bar{n}_1 s_{10}/2 + r_2 n_2 \bar{n}_2 s_4],$$

$$\Pi_5(t) = \lambda[-r_1 n_1 m_1 s_2 + r_1 n_1 \ell s_5 + r_1(1 + r_1 r_2) n_1 m_2 s_{14}/2 - r_1(1 + r_1 r_2) n_1 \ell s_{12}/2$$
$$+ r_1(1 + r_1 r_2) n_2 m_1 s_{10}/2 - r_1(1 + r_1 r_2) n_2 \ell s_8/2 - r_2 r_1^2 n_2 m_2 s_4 + r_2 r_1^2 n_2 \ell s_6],$$
$$\Pi_6(t) = \lambda[r_1 r_2 n_1 m_1 s_2 - r_1 r_2 n_1 \ell s_5 - (1 + r_1 r_2) n_1 m_2 s_{14}/2 + (1 + r_1 r_2) n_1 \ell s_{12}/2 \quad \text{(A18)}$$
$$- r_1 r_2(1 + r_1 r_2) n_2 m_1 s_{10}/2 + r_1 r_2(1 + r_1 r_2) n_2 \ell s_8/2 + r_1 r_2 n_2 m_2 s_4 - r_1 r_2 n_2 \ell s_6],$$

$$\Pi_7(t) = \lambda[r_1 r_2 n_1 m_1 s_2 - r_1 r_2 n_1 \ell s_5 - r_1 r_2(1 + r_1 r_2) n_1 m_2 s_{14}/2 + r_1 r_2(1 + r_1 r_2) n_1 \ell s_{12}/2$$
$$- (1 + r_1 r_2) n_2 m_1 s_{10}/2 + (1 + r_1 r_2) n_2 \ell s_8/2 + r_1 r_2 n_2 m_2 s_4 - r_1 r_2 n_2 \ell s_6],$$
$$\Pi_8(t) = \lambda[-r_1 r_2^2 n_1 m_1 s_2 + r_1 r_2^2 n_1 \ell s_5 + r_2(1 + r_1 r_2) n_1 m_2 s_{14}/2 - r_2(1 + r_1 r_2) n_1 \ell s_{12}/2 \quad \text{(A19)}$$
$$+ r_2(1 + r_1 r_2) n_2 m_1 s_{10}/2 - r_2(1 + r_1 r_2) n_2 \ell s_8/2 - r_2 n_2 m_2 s_4 + r_2 n_2 \ell s_6],$$

$$\Pi_9(t) = \lambda[-r_1 m_1 \bar{n}_1 s_2 + r_1(1+r_1 r_2) m_1 \bar{n}_2 s_{14}/2 + r_1 \bar{n}_1 \ell s_5 - r_1(1+r_1 r_2) \bar{n}_2 \ell s_{13}/2$$
$$+ r_1(1+r_1 r_2) m_2 \bar{n}_1 s_{10}/2 - r_2 r_1^2 m_2 \bar{n}_2 s_4 - r_2(1+r_1 r_2) \bar{n}_1 \ell s_9/2 + r_2 r_1^2 \bar{n}_2 \ell s_6],$$

$$\Pi_{10}(t) = \lambda[r_1 r_2 m_1 \bar{n}_1 s_2 - (1+r_1 r_2) m_1 \bar{n}_2 s_{14}/2 - r_1 r_2 \bar{n}_1 \ell s_5 + (1+r_1 r_2) \bar{n}_2 \ell s_{13}/2$$
$$- r_1 r_2(1+r_1 r_2) m_2 \bar{n}_1 s_{10}/2 + r_1 r_1^2 m_2 \bar{n}_2 s_4 + r_1 r_2(1+r_1 r_2) \bar{n}_1 \ell s_9/2 - r_1 r_2 \bar{n}_2 \ell s_6],$$
(A20)

$$\Pi_{11}(t) = \lambda[r_1 r_2 m_1 \bar{n}_1 s_2 - r_1 r_2(1+r_1 r_2) m_1 \bar{n}_2 s_{14}/2 - r_1 r_2 \bar{n}_1 \ell s_5 + r_1 r_2(1+r_1 r_2) \bar{n}_2 \ell s_{13}/2$$
$$-(1+r_1 r_2) m_2 \bar{n}_1 s_{10}/2 + r_1 r_2 m_2 \bar{n}_2 s_4 + (1+r_1 r_2) \bar{n}_1 \ell s_9/2 - r_1 r_2 \bar{n}_2 \ell s_6],$$

$$\Pi_{12}(t) = \lambda[-r_1 r_2^2 m_1 \bar{n}_1 s_2 + r_2(1+r_1 r_2) m_1 \bar{n}_2 s_{14}/2 + r_1 r_2^2 \bar{n}_1 \ell s_5 - r_2(1+r_1 r_2) \bar{n}_2 \ell s_{13}/2$$
$$+ r_2(1+r_1 r_2) m_2 \bar{n}_1 s_{10}/2 - r_2 m_2 \bar{n}_2 s_4 - r_2(1+r_1 r_2) \bar{n}_1 \ell s_9/2 + r_2 \bar{n}_2 \ell s_6],$$
(A21)

$$\Pi_{13}(t) = \lambda[r_1 m_1^2 s_2 - 2 r_1 m_1 \ell s_5 - r_1(1+r_1 r_2) m_1 m_2 s_{14}/2 + r_1(1+r_1 r_2) m_1 \ell s_{12}/2$$
$$+ r_1 \ell^2 s_1 + r_1(1+r_1 r_2) m_2 \ell s_{13}/2 - r_1(1+r_1 r_2) \ell^2 s_{11}/2 - r_1(1+r_1 r_2) m_1 m_2 s_{10}/2 + r_1(1+r_1 r_2) m_2 \ell s_8/2$$
$$+ r_2 r_1^2 m_2^2 s_4 - 2 r_2 r_1^2 m_2 \ell s_6 + r_1(1+r_1 r_2) m_1 \ell s_9/2 - r_1(1+r_1 r_2) \ell^2 s_7/2 + r_2 r_1^2 \ell^2 s_3],$$

$$\Pi_{14}(t) = \lambda[-r_1 r_2 m_1^2 s_2 + 2 r_1 r_2 m_1 \ell s_5 + (1+r_1 r_2) m_1 m_2 s_{14}/2 - (1+r_1 r_2) m_1 \ell s_{12}/2$$
$$- r_1 r_2 \ell^2 s_1 - (1+r_1 r_2) m_2 \ell s_{13}/2 + (1+r_1 r_2) \ell^2 s_{11}/2 + r_1 r_2(1+r_1 r_2) m_1 m_2 s_{10}/2 - r_1 r_2(1+r_1 r_2) m_2 \ell s_8/2$$
$$- r_1 r_2 m_2^2 s_4 + 2 r_1 r_2 m_2 \ell s_6 - r_1 r_2(1+r_1 r_2) m_1 \ell s_9/2 + r_1 r_2(1+r_1 r_2) \ell^2 s_7/2 - r_1 r_2 \ell^2 s_3],$$
(A22)

$$\Pi_{15}(t) = \lambda[-r_1 r_2 m_1^2 s_2 + 2 r_1 r_2 m_1 \ell s_5 + r_1 r_2(1+r_1 r_2) m_1 m_2 s_{14}/2 - r_1 r_2(1+r_1 r_2) m_1 \ell s_{12}/2$$
$$- r_1 r_2 \ell^2 s_1 - r_1 r_2(1+r_1 r_2) m_2 \ell s_{13}/2 + r_1 r_2(1+r_1 r_2) \ell^2 s_{11}/2 + (1+r_1 r_2) m_1 m_2 s_{10}/2 - (1+r_1 r_2) m_2 \ell s_8/2$$
$$- r_1 r_2 m_2^2 s_4 + 2 r_1 r_2 m_2 \ell s_6 - (1+r_1 r_2) m_1 \ell s_9/2 + (1+r_1 r_2) \ell^2 s_7/2 - r_1 r_2 \ell^2 s_3],$$

$$\Pi_{16}(t) = \lambda[r_1 r_2^2 m_1^2 s_2 - 2 r_1 r_2^2 m_1 \ell s_5 - r_2(1+r_1 r_2) m_1 m_2 s_{14}/2 + r_2(1+r_1 r_2) m_1 \ell s_{12}/2$$
$$+ r_1 r_2^2 \ell^2 s_1 + r_2(1+r_1 r_2) m_2 \ell s_{13}/2 - r_2(1+r_1 r_2) \ell^2 s_{11}/2 - r_2(1+r_1 r_2) m_1 m_2 s_{10}/2 + r_2(1+r_1 r_2) m_2 \ell s_8/2$$
$$+ r_2 m_2^2 s_4 - 2 r_2 m_2 \ell s_6 + r_2(1+r_1 r_2) m_1 \ell s_9/2 - r_2(1+r_1 r_2) \ell^2 s_7/2 + r_2 \ell^2 s_3],$$
(A23)

where $s_k \equiv s_k(t)$, $(k=1,...,14)$ and

$$s_1(t) = t/2 + \sin(2\Omega_1 t)/4\Omega_1, \quad s_2(t) = t/2 - \sin(2\Omega_1 t)/4\Omega_1, \quad s_3(t) = t/2 + \sin(2\Omega_2 t)/4\Omega_2,$$
$$s_4(t) = t/2 - \sin(2\Omega_2 t)/4\Omega_2, \quad s_5(t) = \sin^2(\Omega_1 t)/2\Omega_1, \quad s_6(t) = \sin^2(\Omega_2 t)/2\Omega_2,$$
(A24)

$$s_7(t) = s_{11}(t) = \frac{\Omega_1 \cos(t\Omega_2) \sin(t\Omega_1) - \Omega_2 \cos(t\Omega_1) \sin(t\Omega_2)}{\Omega_1^2 - \Omega_2^2},$$
(A25)

$$s_8(t) = s_{13}(t) = \frac{-\Omega_2 + \Omega_2 \cos(t\Omega_2) \cos(t\Omega_1) + \Omega_1 \sin(t\Omega_1) \sin(t\Omega_2)}{\Omega_1^2 - \Omega_2^2},$$
(A26)

$$s_9(t) = s_{12}(t) = \frac{\Omega_1 - \Omega_1 \cos(t\Omega_2) \cos(t\Omega_1) - \Omega_2 \sin(t\Omega_1) \sin(t\Omega_2)}{\Omega_1^2 - \Omega_2^2},$$
(A27)

$$s_{10}(t) = s_{14}(t) = \frac{\Omega_2 \cos(t\Omega_2) \sin(t\Omega_1) - \Omega_1 \cos(t\Omega_1) \sin(t\Omega_2)}{\Omega_1^2 - \Omega_2^2}.$$
(A28)

The functions $A_{1,2}, B_{1,2}, C_{1,2}, E_{1,2,3,4}$ in Eq.(6) designated as $R_k$ ($k=1,2$) can be represented as follows

$$R_k(t) = \frac{2M\gamma}{\pi}\left\{\int_0^\infty d\omega\,\omega\,\mathrm{Coth}\left(\frac{\hbar\omega}{2k_B T_1}\right)\int_0^t\int_0^\tau ds\,d\tau\, R_k^{(1)}(\tau,s)\cos[\omega(\tau-s)]\exp[\gamma(\tau+s)]\right.$$
$$\left.+\int_0^\infty d\omega\,\omega\,\mathrm{Coth}\left(\frac{\hbar\omega}{2k_B T_2}\right)\int_0^t\int_0^\tau ds\,d\tau\, R_k^{(2)}(\tau,s)\cos[\omega(\tau-s)]\exp[\gamma(\tau+s)]\right\} \quad (A29)$$

where

$$A_1^{(1)}(\tau,s) = \bar{n}_1^2 f_1 - r_1 r_2 \bar{n}_1 \bar{n}_2 (f_3+f_4) + r_1^2 r_2^2 \bar{n}_2^2 f_2,\quad A_1^{(2)}(\tau,s) = r_1^2[\bar{n}_1^2 f_1 - \bar{n}_1\bar{n}_2(f_3+f_4) + \bar{n}_2^2 f_2],$$
$$A_2^{(1)}(\tau,s) = r_2^2[\bar{n}_1^2 f_1 - \bar{n}_1\bar{n}_2(f_3+f_4) + \bar{n}_2^2 f_2],\quad A_2^{(2)}(\tau,s) = \bar{n}_2^2 f_2 - r_1 r_2 \bar{n}_1 \bar{n}_2 (f_3+f_4) + r_1^2 r_2^2 \bar{n}_1^2 f_1, \quad (A30)$$

$$B_1^{(1)}(\tau,s) = \tilde{q}_1(\tau,s) + r_1 r_2 \tilde{d}(\tau,s) + r_1^2 r_2^2 \tilde{q}_2(\tau,s),\quad B_1^{(2)}(\tau,s) = r_1^2[\tilde{q}_1(\tau,s) + \tilde{d}(\tau,s) + \tilde{q}_2(\tau,s)],$$
$$B_2^{(1)}(\tau,s) = r_2^2[\tilde{q}_1(\tau,s) + \tilde{d}(\tau,s) + \tilde{q}_2(\tau,s)],\quad B_2^{(2)}(\tau,s) = \tilde{q}_2(\tau,s) + r_1 r_2 \tilde{d}(\tau,s) + r_1^2 r_2^2 \tilde{q}_1(\tau,s), \quad (A31)$$

where

$$\tilde{q}_1(\tau,s) = -2\bar{n}_1 m_1 f_1 + \bar{n}_1 \ell(f_9 + f_{10}),\quad \tilde{q}_2(\tau,s) = -2\bar{n}_2 m_2 f_2 + \bar{n}_2 \ell(f_{13} + f_{14}),$$
$$\tilde{d}(\tau,s) = (\bar{n}_2 m_1 + \bar{n}_1 m_2)(f_3+f_4) - \bar{n}_1 \ell(f_{11} + f_{16}) - \bar{n}_2 \ell(f_{12} + f_{15}), \quad (A32)$$

$$C_1^{(1)}(\tau,s) = q_1(\tau,s) + r_1 r_2 d(\tau,s) + r_1^2 r_2^2 q_2(\tau,s),\quad C_1^{(2)}(\tau,s) = r_1^2[q_1(\tau,s) + d(\tau,s) + q_2(\tau,s)],$$
$$C_2^{(1)}(\tau,s) = r_2^2[q_1(\tau,s) + d(\tau,s) + q_2(\tau,s)],\quad C_2^{(2)}(\tau,s) = q_2(\tau,s) + r_1 r_2 d(\tau,s) + r_1^2 r_2^2 q_1(\tau,s), \quad (A33)$$

$$E_1^{(1)}(\tau,s) = -r_2[2q_1(\tau,s) + 2r_1 r_2 q_2(\tau,s) + (1+r_1 r_2)d(\tau,s)],$$
$$E_1^{(2)}(\tau,s) = -r_1[2q_2(\tau,s) + 2r_1 r_2 q_1(\tau,s) + (1+r_1 r_2)d(\tau,s)], \quad (A34)$$

where

$$q_1(\tau,s) = m_1^2 f_1 + \ell^2 f_5 - m_1 \ell(f_9 + f_{10}),\quad q_2(\tau,s) = m_2^2 f_2 + \ell^2 f_6 - m_2 \ell(f_{13}+f_{14}),$$
$$d(\tau,s) = -m_1 m_2(f_3+f_4) - \ell^2(f_7+f_8) + m_1 \ell(f_{11}+f_{16}) + m_2 \ell(f_{12}+f_{15}), \quad (A35)$$

$$E_2^{(1)}(\tau,s) = r_2\{q'(\tau,s) + d'(\tau,s) + r_1 r_2[q''(\tau,s) + d''(\tau,s)]\},$$
$$E_2^{(2)}(\tau,s) = r_1\{q''(\tau,s) + d'(\tau,s) + r_1 r_2[q'(\tau,s) + d''(\tau,s)]\}, \quad (A36)$$

$$E_3^{(1)}(\tau,s) = r_2\{q'(\tau,s) + d''(\tau,s) + r_1 r_2[q''(\tau,s) + d'(\tau,s)]\},$$
$$E_3^{(2)}(\tau,s) = r_1\{q''(\tau,s) + d''(\tau,s) + r_1 r_2[q'(\tau,s) + d'(\tau,s)]\}, \quad (A37)$$

where

$$q'(\tau,s) = 2\bar{n}_1 m_1 f_1 - \bar{n}_1 \ell(f_9 + f_{10}),\quad d'(\tau,s) = \bar{n}_2 \ell(f_{12}+f_{15}) - \bar{n}_2 m_1(f_3+f_4),$$
$$q''(\tau,s) = 2\bar{n}_2 m_2 f_2 - \bar{n}_2 \ell(f_{13}+f_{14}),\quad d''(\tau,s) = \bar{n}_1 \ell(f_{11}+f_{16}) - \bar{n}_1 m_2(f_3+f_4), \quad (A38)$$

$$E_4^{(1)}(\tau,s) = r_2\{\bar{n}_1 \bar{n}_2(f_3+f_4) - 2\bar{n}_1^2 f_1 + r_1 r_2[\bar{n}_1 \bar{n}_2(f_3+f_4) - \bar{n}_2^2 f_2]\},$$
$$E_4^{(2)}(\tau,s) = r_1\{\bar{n}_1 \bar{n}_2(f_3+f_4) - 2\bar{n}_2^2 f_2 + r_1 r_2[\bar{n}_1 \bar{n}_2(f_3+f_4) - \bar{n}_1^2 f_1]\}, \quad (A39)$$

where all $f_i \equiv f_i(\tau, s)$ $(i = 1,...,16)$ are the functions of $\tau$ and $s$:

$$
\begin{aligned}
&f_1(\tau,s) = sin(\Omega_1\tau)sin(\Omega_1 s), \quad f_2(\tau,s) = sin(\Omega_2\tau)sin(\Omega_2 s), \quad f_3(\tau,s) = sin(\Omega_1\tau)sin(\Omega_2 s), \\
&f_4(\tau,s) = sin(\Omega_2\tau)sin(\Omega_1 s), \quad f_5(\tau,s) = cos(\Omega_1\tau)cos(\Omega_1 s), \quad f_6(\tau,s) = cos(\Omega_2\tau)cos(\Omega_2 s), \\
&f_7(\tau,s) = cos(\Omega_1\tau)cos(\Omega_2 s), \quad f_8(\tau,s) = cos(\Omega_2\tau)cos(\Omega_1 s), \quad f_9(\tau,s) = sin(\Omega_1\tau)cos(\Omega_1 s), \\
&f_{10}(\tau,s) = cos(\Omega_1\tau)sin(\Omega_1 s), \quad f_{11}(\tau,s) = sin(\Omega_1\tau)cos(\Omega_2 s), \quad f_{12}(\tau,s) = cos(\Omega_1\tau)sin(\Omega_2 s), \\
&f_{13}(\tau,s) = sin(\Omega_2\tau)cos(\Omega_2 s), \quad f_{14}(\tau,s) = cos(\Omega_2\tau)sin(\Omega_1 s), \quad f_{15}(\tau,s) = sin(\Omega_2\tau)cos(\Omega_1 s), \\
&f_{16}(\tau,s) = cos(\Omega_2\tau)sin(\Omega_1 s).
\end{aligned}
\quad (A40)
$$

We underline that the integration with respect to $\tau$ and $s$ in Eqs.(A29) can be performed analytically, but the final results are very cumbersome and we wrote down here only integral forms.

### Appendix B. Solution to the motion equations for classical paths.

We obtained in our paper [46] two pairs of coupled equations for classical paths

$$
\begin{cases} \ddot{\tilde{X}}_1 + 2\gamma_1 \dot{\tilde{X}}_1 + \omega_{01}^2 \tilde{X}_1 = (\lambda/M_1)\tilde{X}_2 \\ \ddot{\tilde{X}}_2 + 2\gamma_2 \dot{\tilde{X}}_2 + \omega_{02}^2 \tilde{X}_2 = (\lambda/M_2)\tilde{X}_1 \end{cases}, \quad (B1)
$$

$$
\begin{cases} \ddot{\tilde{\xi}}_1 - 2\gamma_1 \dot{\tilde{\xi}}_1 + \omega_{01}^2 \tilde{\xi}_1 = (\lambda/M_1)\tilde{\xi}_2 \\ \ddot{\tilde{\xi}}_2 - 2\gamma_2 \dot{\tilde{\xi}}_2 + \omega_{02}^2 \tilde{\xi}_2 = (\lambda/M_2)\tilde{\xi}_1. \end{cases} \quad (B2)
$$

Then, we seek a solution of the form $\tilde{X}_{1,2} = A_{1,2} \exp(\varepsilon\tau)$ taking $\varepsilon = i\omega - \delta$. From Eq.(B1) we obtain a determinant equation, yielding for the real and imaginary parts the pair of equations

$$
\omega^4 - (\omega_{01}^2 + \omega_{02}^2 + \Delta_1)\omega^2 - \lambda^2/M_1 M_2 + \Delta_2 = 0, \quad (B3)
$$

$$
\begin{aligned}
&(\omega^2 - \omega_{02}^2)(\gamma_1 - \delta) + (\omega^2 - \omega_{01}^2)(\gamma_2 - \delta) \\
&-(\gamma_1 - \delta)(2\gamma_2\delta - \delta^2) - (\gamma_2 - \delta)(2\gamma_1\delta - \delta^2) = 0
\end{aligned}, \quad (B4)
$$

where $\Delta_1 = 4(\gamma_1 - \delta)(\gamma_2 - \delta) - (2\gamma_1\delta - \delta^2) - (2\gamma_2\delta - \delta^2)$,

$\Delta_2 = \omega_{01}^2\omega_{02}^2 + (2\gamma_1\delta - \delta^2)(2\gamma_2\delta - \delta^2) - (2\gamma_1\delta - \delta^2)\omega_{02}^2 - (2\gamma_2\delta - \delta^2)\omega_{01}^2$.

We can see from Eq.(B4) that in case of $\gamma_1 = \gamma_2 = \gamma$ we have the simpler equation

$$
(\gamma - \delta)[(\omega^2 - \omega_{02}^2) + (\omega^2 - \omega_{01}^2) - 2(2\gamma\delta - \delta^2)] = 0, \quad (B5)
$$

where from it is clear that the Eq.(B4) can be satisfied by $\delta = \gamma$.

Here, we would like to emphasize that the condition $\gamma_1 = \gamma_2 = \gamma$ is a single restriction in our work, but this restriction essentially simplifies the procedure of solving the problem in a general case. The roots of the equation (B3) in this case are

$$\Omega_{1,2}^2 = (\omega_{01}^2 + \omega_{02}^2 - 2\gamma^2)/2 \mp \sqrt{(\omega_{02}^2 - \omega_{01}^2)^2/4 + \lambda^2/M_1 M_2}. \tag{B6}$$

where we ordered the new eigenfrequencies as follows $\Omega_1 < \omega_{01} < \omega_{02} < \Omega_2$.

After substitution of $\Omega_{1,2}^2$ into Eq.(B1) in the form $\tilde{X}_{1,2} = A_{1,2} \exp(i\Omega_{1,2}\tau - \gamma\tau)$ we obtain two ratios

$$r_1 \equiv r_1(\Omega_1) = \frac{A_2}{A_1} = \frac{(\omega_{01}^2 - \gamma^2) - \Omega_1^2}{\lambda/M_1},$$

$$r_2 \equiv r_2(\Omega_2) = \frac{A_1}{A_2} = \frac{(\omega_{02}^2 - \gamma^2) - \Omega_2^2}{\lambda/M_2}, \tag{B7}$$

which determine a relative contribution to the oscillators vibrations from the first or from the second eigenmodes, correspondingly. It is clearly seen from a general solution

$$\tilde{X}_1(\tau) = B_1 \sin(\Omega_1 \tau + \varphi_1)\exp(-\delta_1 \tau) + r_2 B_2 \sin(\Omega_2 \tau + \varphi_2)\exp(-\delta_2 \tau),$$

$$\tilde{X}_2(\tau) = r_1 B_1 \sin(\Omega_1 \tau + \varphi_1)\exp(-\delta_1 \tau) + B_2 \sin(\Omega_2 \tau + \varphi_2)\exp(-\delta_2 \tau). \tag{B8}$$

The above equations take into account the contributions from the first and second modes to the classical trajectory $\tilde{X}_1$ and $\tilde{X}_2$, see for instance [53-55].

Satisfying the conditions $X_{1,2}(0) = X_{i1,2}$ and $X_{1,2}(t) = X_{f1,2}$, we obtain

$$B_1 \cos\varphi_1 = \frac{X_{f1} - r_2 X_{f2}}{(1 - r_1 r_2)\sin(\Omega_1 t)} \exp(\delta_1 t) - \cot(\Omega_1 t) \frac{X_{i1} - r_2 X_{i2}}{(1 - r_1 r_2)},$$

$$B_2 \cos\varphi_2 = \frac{X_{f2} - r_1 X_{f1}}{(1 - r_1 r_2)\sin(\Omega_2 t)} \exp(\delta_2 t) - \cot(\Omega_2 t) \frac{X_{i2} - r_1 X_{i1}}{(1 - r_1 r_2)}, \tag{B9}$$

$$B_1 \sin\varphi_1 = \frac{X_{i1} - r_2 X_{i2}}{(1 - r_1 r_2)}, \quad B_2 \sin\varphi_2 = \frac{X_{i2} - r_1 X_{i1}}{(1 - r_1 r_2)} \tag{B10}$$

for Eq.(B8).

In its turn, equations (B2) for the classical paths $\tilde{\xi}_{1,2}(\tau)$ of backward amplitudes can be solved by analogy. Corresponding formulas are listed in Eqs.(A10),(A11) in [46].

**Appendix C. Time dependent functions in Eq.(11)**

Below we keep in mind that $\rho_0(t) \equiv \rho_0(t, \lambda, T_1, T_2)$, $g(t) \equiv g(t, \lambda, T_1, T_2)$, $g'(t) \equiv g'(t, \lambda, T_1, T_2)$, $g''(t) \equiv g''(t, \lambda, T_1, T_2)$ and we have

$$\rho_0(t) = \frac{F_1^2(t) F_2^2(t)}{\sqrt{\sigma_{01}^2 \sigma_{02}^2}} \frac{\pi \hbar \tilde{C}_1 \tilde{C}_2}{\sqrt{Z_1 Y_1 [4\hbar a_2 (C_2 + \hbar a_2) + (D'_4 + \Pi_{16})^2]}}, \tag{C1}$$

where $a_1 = 1/8\sigma_{01}^2$, $a_2 = 1/8\sigma_{02}^2$, $\tilde{C}_{1,2}$ are the normalized constants, and $F(t)$ is the wave function amplitude for the undamped case with renormalized eigenfrequencies of two oscillators due to their coupling, see, please Appendix D in [46],

$$g_1(t) = \left[\frac{(D_9 + D_9' + \Pi_6)^2}{4\hbar(C_2 + \hbar a_2)} - \frac{e_6^2(C_2/\hbar + a_2)}{4\hbar a_2(C_2 + \hbar a_2) + (D_4' + \Pi_{16})^2} + \frac{Z_2^2}{4\hbar^2 Z_1} + \frac{Y_5^2}{4\hbar^2 Y_1}\right],$$

$$g_{12}(t) = \left[\frac{2(D_3' + \Pi_8)(D_9 + D_9' + \Pi_6)}{4\hbar(C_2 + \hbar a_2)} - \frac{2e_5 e_6(C_2/\hbar + a_2)}{4\hbar a_2(C_2 + \hbar a_2) + (D_4' + \Pi_{16})^2} + \frac{Z_2 Z_3}{2\hbar^2 Z_1} + \frac{Y_4 Y_5}{2\hbar^2 Y_1}\right], \quad (C2)$$

$$g_2(t) = \left[\frac{(D_3' + \Pi_8)^2}{4\hbar(C_2 + \hbar a_2)} - \frac{e_5^2(C_2/\hbar + a_2)}{4\hbar a_2(C_2 + \hbar a_2) + (D_4' + \Pi_{16})^2} + \frac{Z_3^2}{4\hbar^2 Z_1} + \frac{Y_4^2}{4\hbar^2 Y_1}\right],$$

$$g_1'(t) = \left[\frac{A_1}{\hbar} - \frac{E_3^2}{4\hbar(C_2 + \hbar a_2)} - \frac{e_2^2(C_2/\hbar + a_2)}{4\hbar a_2(C_2 + \hbar a_2) + (D_4' + \Pi_{16})^2} + \frac{Z_4^2}{4\hbar^2 Z_1} + \frac{Y_2^2}{4\hbar^2 Y_1}\right],$$

$$g_{12}'(t) = \left[\frac{E_4}{\hbar} - \frac{B_2 E_3}{2\hbar(C_2 + \hbar a_2)} - \frac{2e_1 e_1(C_2/\hbar + a_2)}{4\hbar a_2(C_2 + \hbar a_2) + (D_4' + \Pi_{16})^2} + \frac{Z_4 Z_5}{2\hbar^2 Z_1} + \frac{Y_2 Y_3}{2\hbar^2 Y_1}\right], \quad (C3)$$

$$g_2'(t) = \left[\frac{A_2}{\hbar} - \frac{B_2^2}{4\hbar(C_2 + \hbar a_2)} - \frac{e_1^2(C_2/\hbar + a_2)}{4\hbar a_2(C_2 + \hbar a_2) + (D_4' + \Pi_{16})^2} + \frac{Z_5^2}{4\hbar^2 Z_1} + \frac{Y_3^2}{4\hbar^2 Y_1}\right],$$

$$g_{11}''(t) = \left[-\frac{i(D_1 + \Pi_1)}{\hbar} + \frac{iE_3(D_9 + D_9' + \Pi_6)}{2\hbar(C_2 + \hbar a_2)} + \frac{2e_2 e_6(C_2/\hbar + a_2)}{4\hbar a_2(C_2 + \hbar a_2) + (D_4' + \Pi_{16})^2} + \frac{Z_4 Z_6}{4\hbar^2 Z_1} + \frac{Y_2 Y_5}{4\hbar^2 Y_1}\right],$$

$$g_{21}''(t) = \left[-\frac{i(D_5 + D_5' + \Pi_3)}{\hbar} + \frac{iE_3(D_3' + \Pi_8)}{2\hbar(C_2 + \hbar a_2)} - \frac{2e_5 e_5(C_2/\hbar + a_2)}{4\hbar a_2(C_2 + \hbar a_2) + (D_4' + \Pi_{16})^2} + \frac{Z_3 Z_4}{2\hbar^2 Z_1} + \frac{Y_2 Y_4}{2\hbar^2 Y_1}\right],$$

$$g_{12}''(t) = \left[-\frac{i(D_6 + D_6' + \Pi_2)}{\hbar} + \frac{iB_2(D_9 + D_9' + \Pi_6)}{2\hbar(C_2 + \hbar a_2)} - \frac{2e_1 e_6(C_2/\hbar + a_2)}{4\hbar a_2(C_2 + \hbar a_2) + (D_4' + \Pi_{16})^2} + \frac{Z_2 Z_5}{2\hbar^2 Z_1}\right. \quad (C4)$$

$$\left.+\frac{Y_3 Y_5}{2\hbar^2 Y_1}\right],$$

$$g_{22}''(t) = \left[-\frac{i(D_1' + \Pi_4)}{\hbar} + \frac{iB_2(D_3' + \Pi_8)}{2\hbar(C_2 + \hbar a_2)} + \frac{2e_1 e_5(C_2/\hbar + a_2)}{4\hbar a_2(C_2 + \hbar a_2) + (D_4' + \Pi_{16})^2} + \frac{Z_3 Z_5}{4\hbar^2 Z_1} + \frac{Y_3 Y_4}{4\hbar^2 Y_1}\right].$$

The time-dependent functions in Eqs.(C1)-(C4) can be found in Appendix A. Besides, in Eqs.(C2)-(C4) for functions $e_k = e_k(t), (k = 1,...,6)$ we have

$$e_1 = D_2' + \Pi_{12} - \frac{B_2(D_4' + \Pi_{16})}{2(C_2 + \hbar a_2)}, \; e_2 = D_8 + D_8' + \Pi_{11} - \frac{E_3(D_4' + \Pi_{16})}{2(C_2 + \hbar a_2)},$$

$$e_3 = D_{12} + D_{12}' + \Pi_{15} - \frac{E_1(D_4' + \Pi_{16})}{2(C_2 + \hbar a_2)}, \; e_4 = \frac{(D_{11} + D_{11}' + \Pi_{14})(D_4' + \Pi_{16})}{2(C_2 + \hbar a_2)}, \quad (C5)$$

$$e_5 = \frac{(D_3' + \Pi_8)(D_4' + \Pi_{16})}{2(C_2 + \hbar a_2)}, \; e_6 = \frac{(D_9 + D_9' + \Pi_6)(D_4' + \Pi_{16})}{2(C_2 + \hbar a_2)},$$

for functions $Z_k \equiv Z_k(t)$, $(k=1,...,6)$

$$Z_1 = C_1/\hbar + a_1 - \frac{E_1^2}{4\hbar(C_2+\hbar a_2)} + \frac{e_3^2(C_2/\hbar+a_2)}{4\hbar a_2(C_2+\hbar a_2)+(D_4'+\Pi_{16})^2},$$

$$Z_2 = D_3 + \Pi_5 - \frac{E_1(D_9+D_9'+\Pi_6)}{2(C_2+\hbar a_2)} - \frac{2e_3 e_6(C_2+\hbar a_2)}{4\hbar a_2(C_2+\hbar a_2)+(D_4'+\Pi_{16})^2},$$

$$Z_3 = D_{10} + D_{10}' + \Pi_7 - \frac{E_1(D_3'+\Pi_8)}{2(C_2+\hbar a_2)} - \frac{2e_3 e_5(C_2+\hbar a_2)}{4\hbar a_2(C_2+\hbar a_2)+(D_4'+\Pi_{16})^2},$$

$$Z_4 = iB_1 + \frac{i2e_2 e_3(C_2+\hbar a_2)}{4\hbar a_2(C_2+\hbar a_2)+(D_4'+\Pi_{16})^2},$$

$$Z_5 = iE_2 - \frac{iE_1(B_2+E_3)}{2(C_2+\hbar a_2)} + \frac{i2e_1 e_3(C_2+\hbar a_2)}{4\hbar a_2(C_2+\hbar a_2)+(D_4'+\Pi_{16})^2},$$

$$Z_6 = D_4 + \Pi_{13} - \frac{E_1(D_{11}+D_{11}'+\Pi_{14})}{2(C_2+\hbar a_2)} - \frac{2e_3 e_4(C_2+\hbar a_2)}{4\hbar a_2(C_2+\hbar a_2)+(D_4'+\Pi_{16})^2},$$

(C6)

for functions $Y_k \equiv Y_k(t)$, $(k=1,...,5)$

$$Y_1 = a_1 + \frac{(D_{11}+D_{11}'+\Pi_{14})^2}{4\hbar(C_2+\hbar a_2)} - \frac{e_4^2(C_2/\hbar+a_2)}{4\hbar a_2(C_2+\hbar a_2)+(D_4'+\Pi_{16})^2} + \frac{Z_6^2}{4\hbar^2 Z_1},$$

$$Y_2 = D_2 + \Pi_9 - \frac{E_3(D_{11}+D_{11}'+\Pi_{14})}{2(C_2+\hbar a_2)} - \frac{2e_2 e_4(C_2+\hbar a_2)}{4\hbar a_2(C_2+\hbar a_2)+(D_4'+\Pi_{16})^2} + \frac{iZ_4 Z_6}{2\hbar Z_1},$$

$$Y_3 = D_7 + D_7' + \Pi_{10} - \frac{B_2(D_{11}+D_{11}'+\Pi_{14})}{2(C_2+\hbar a_2)} - \frac{2e_1 e_4(C_2+\hbar a_2)}{4\hbar a_2(C_2+\hbar a_2)+(D_4'+\Pi_{16})^2} + \frac{iZ_5 Z_6}{2\hbar Z_1},$$

$$Y_4 = \frac{i(D_3'+\Pi_8)(D_{11}+D_{11}'+\Pi_{14})}{2(C_2+\hbar a_2)} - \frac{i2e_4 e_5(C_2+\hbar a_2)}{4\hbar a_2(C_2+\hbar a_2)+(D_4'+\Pi_{16})^2} + \frac{iZ_3 Z_6}{2\hbar Z_1},$$

$$Y_5 = \frac{i(D_9+D_9'+\Pi_6)(D_{11}+D_{11}'+\Pi_{14})}{2(C_2+\hbar a_2)} - \frac{i2e_4 e_6(C_2+\hbar a_2)}{4\hbar a_2(C_2+\hbar a_2)+(D_4'+\Pi_{16})^2} + \frac{iZ_2 Z_6}{2\hbar Z_1}.$$

(C7)